\documentclass[%
	rsi,
	amsmath,amssymb,
	 preprint,%
	]{revtex4-2}
	\usepackage[pdftex]{graphicx}
	\graphicspath{{media/}}
	\usepackage[x11names,table]{xcolor} %Para tablas con colores
	\usepackage{lineno}
	\draft % marks overfull lines with a black rule on the right

\begin{document}

	%\preprint{}

	%\title{Design and Efficient Implementation of an Oversampled GDFT\\ Filter Bank
	%Channelizer on a Zynq Device}
	%\title{Implementation of an Oversampled Generalized Discrete Fourier Transform
	%Filter Bank Channelizer for Cryogenic Detectors}
	\title{Oversampled Filter Bank Channelizer for Cryogenic Detectors}



	\author{L. H. Arnaldi}
	\email[]{arnaldi@cab.cnea.gov.ar}%
	%\thanks{}
	%\altaffiliation{Particle and Radiation Laboratory}
	\affiliation{Bariloche Atomic Center and Instituto Balseiro, National
	Commission of Atomic Energy\\ Av. Bustillo 9500 (8400), S. C. de Bariloche,
	Rio Negro, Argentina}

	\author{D. Dellavale}
	%\email[]{}
	%\thanks{}
	\affiliation{National Council for Scientific and Technical Research of Argentina
	and Instituto Balseiro\\ Av. Bustillo 9500 (8400), S. C. de Bariloche,
	Rio Negro, Argentina}
	%\altaffiliation[Also at ]{National Council for Scientific and Technical Research of
	%Argentina}%

	%\author{J. Lipovetzky}
	%\email[]{}
	%\thanks{}
	%\altaffiliation[Also at ]{National Council for Scientific and Technical Research of
	%Argentina}%
	%\affiliation{Bariloche Atomic Center and Instituto Balseiro, National
	%Commission of Atomic Energy}%
	%\affiliation{National Council for Scientific and Technical Research of
	%Argentina, Bariloche Atomic Center and Instituto Balseiro, National
	%Commission of Atomic Energy}


	\date{}   % arXiv: no build date on a 2021 paper

	\begin{abstract}
		We present the field programmable gate array (FPGA) implementation of an
		oversampled generalized DFT	(GDFT) filter bank, optimized for its usage in
		the   readout of cryogenic detectors, such as the microwave kinetic
		inductance detectors (MKID). A polyphase structure is derived underlining
		the efficiency of complex valued subband processing, raised by a GDFT
		modulation. For the latter, a fast Fourier algorithm	realization may be
		used, yielding a highly efficient polyphase implementation for arbitrary
		integer decimation ratios.  The filter bank is tested through a 16-channel
		GDFT channelizer. The implementation is based on the Xilinx Zynq family of
		FPGAs and aims to show the data reuse and flexibility offered by
		the GDFT channelizer structure.  General design criteria are summarized for
		the 16-channel polyphase filter bank channelizer.  The performance in
		hardware resource usage is also presented, taking into account one of the
		main objetives of the current physics experiments, aiming to have extremely
		large readout format.
	\end{abstract}

	%% Note that keywords are not normally used for peerreview papers.
	%\begin{IEEEkeywords}
	%Multirate signal processing, Polyphase Filter Bank, Generalized Discrete Fourier Transform,
	%				FIR Filtering. 
	%\end{IEEEkeywords}

	\maketitle %\maketitle must follow title, authors, abstract and \pacs

	% AIP Publishing credit line for the accepted manuscript after publication (AIP "Sharing Content Online")
	\noindent{\small\textit{This article may be downloaded for personal use only. Any other use requires prior permission of
	the author and AIP Publishing. This article appeared in L. H. Arnaldi and H. D. Dellavale, Rev. Sci. Instrum. 92, 023304
	(2021) and may be found at https://doi.org/10.1063/5.0035449.}}

	% References should be done using the \cite and \label commands
	\section{Introduction}\label{intro}
	Within the particle physics and astrophysics communities,
	cryogenic detectors have been widely used for more than three decades in
	applications that require exceptionally high levels of performance. In this
	group of detectors, superconducting micro-resonators, like the microwave kinetic
	inductance detectors (MKID), especially stand out for being simple and versatile
	devices~\cite{Day2003,Mazin2006a}. Interest in these devices has grown
	especially fast over the last decade as a result of the demonstration of
	transmission line micro-resonators as sensitive photon detectors~\cite{Day2003}
	and the subsequent adoption of this type of resonator for reading quantum state
	of superconducting qubits~\cite{AndreasWallraff2004}.  At present,
	superconducting micro-resonators are under intensive study and are intended for
	a wide variety of applications, using frequency-multiplexed readouts of
	cryogenic detector arrays~\cite{McHugh2012,Henderson2018,VanRantwijk2016}. 

	In radio-astronomy, the current challenge is to manufacture detector matrices
	for image studies in extremely large format, to achieve wide fields of view in
	the instruments and place image matrices in space. Associated, is the challenge
	of achieving the manufacture of electronic excitation and readout systems
	capable of working together with these large detector arrays. These electronic
	systems must have a large computational capacity and be able to process large
	amounts of data in relatively short times. Normally, some kind of filter-bank
	arrangement is used for readout these large detector
	arrays~\cite{Mazin2006,McHugh2012,Mazin2013,VanRantwijk2016}.  
	%de mi intro cap5
	In these demanding digital signal processing (DSP) applications, the received
	signal is oversampled by several orders of magnitude over the required
	sampling rate (Nyquist) for the application. This is a typical case for
	reading superconducting detector arrays, where the bandwidth of the individual
	resonators ($\sim 40-60$\,kHz) is much less than the bandwidth of the
	frequency division multiplexed (FDM) signal ($\sim 2-8$\,GHz) transmitted
	throughout the array~\cite{McHugh2012,Mazin2013,VanRantwijk2016}. It is also
	the case for highly-multiplexed microwave SQUID~\footnote{Superconducting
	Quantum Interference Devices} readout systems used for Cosmic Microwave
	Background (CMB) and sub-millimeter
	surveys~\cite{Kernasovskiy2018,Henderson2018} and for quantum
	circuits~\cite{Palacios-Laloy2008a}.

	Decimating the signal in a single stage can result in a very expensive and
	computationally complex filtering arrangement. For these cases, implementing the
	sampling rate conversion in multiple decimation stages may be more
	efficient~\cite{Herrmann1973,Rabiner1973}. The gradual decrease in the sampling
	rate results in a simplified design of the various filter stages and the
	resulting design can still meet the general requirements for decimation and
	filtering.  Savings are achieved by reducing the number of multiplications and
	additions per second (MADS), or equivalently, reducing the multiplication
	rate ($R_T$).

	\begin{figure*}[!ht]
		\centering
		\includegraphics[width=0.9\textwidth]{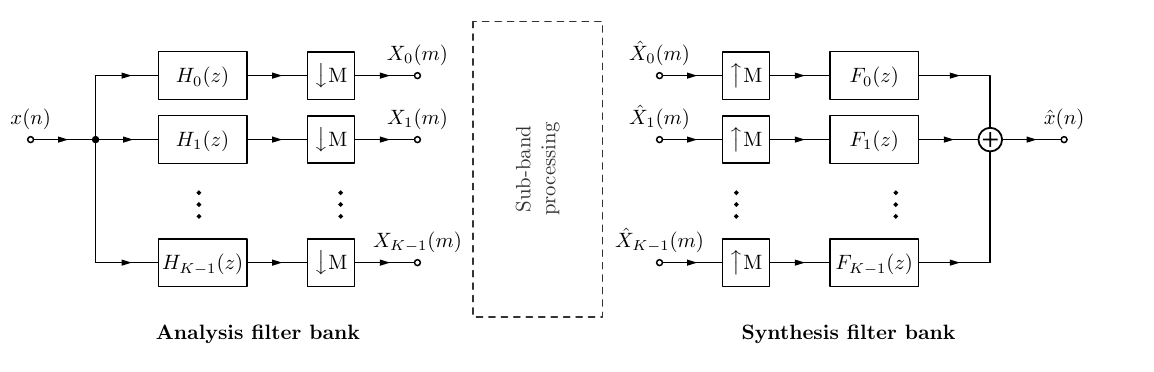}
		%\includegraphics[width=\textwidth]{c2_pfb_ana_synth_bis}
		%\includegraphics[width=0.5\textwidth]{c2_pfb_ana_synth}
		%\caption{Bancos de filtros digitales de$K$canales.}
		\caption{
			Structure of a $K$-channel analysis and synthesis filter bank. The
			$x(n)$ signal is the FDM signal containing the channels of interest.
			The combination of delays, plus down-sampling by $M$ ($\downarrow
			\negmedspace M$) can be seen as a commutator, delivering one input
			sample to each of the filters ($H_k(z)$) at a time and at a low
			sampling rate ($m$).  In the analysis filter bank, every output
			channel $X_k(m)$ is recovered at the output of a M-point DFT block.
			Here, the $n$ and $m$ (with $n >	m$) indexes denote a signal working
			at the input and output sampling rate, respectively. In the synthesis,
			the inverse process occurs. 
			}
			\label{fig_pfb_ana_synth_bis}
	\end{figure*}
	%Optimal solutions to the problem of finding minimum cost functions of multistage
	%structures can be derived analytically~\cite{Coffey2003,Coffey2007}, however
	%these solutions are often groups of non-integer real numbers that cannot be
	%implemented in practical systems. In general, manual adjustment of the results
	%is needed to find the optimal solution set.

	In this paper we focus in the development and implementation of efficient filter
	banks (FB), to extract a set of frequency division multiplexed channels
	that exist in a single sampled data stream. This data stream can be, for
	instance, the result of passing the FDM signal through a matrix of
	superconducting micro-resonators. The filter bank will act as a monitoring and
	analysis system for the superconducting detectors.  Specifically, we consider
	the design of oversampled generalized discrete Fourier transform filter banks
	(GDFT-FB-OVS)~\cite{RonaldE.Crochiere527}, using finite impulse response (FIR)
	filters in its polyphase form as decimators.  This implementation is targeted to
	new FPGA technologies, using standard language libraries and shows to be
	extremely efficient in the use of resources. It differs from previous
	works~\cite{McHugh2012,Mazin2013,VanRantwijk2016} in that it is implemented with
	a more general algorithm (GDFT), giving extra flexibility in the use of the
	available spectrum, maintaining the efficient resource usage. This efficiency is
	given by the use of algorithms such as the FFT and hardware implementations such
	as the polyphase filtering.
	%filters in its polyphase form as multistage decimators.  In this sense, the
	%filters in its polyphase form as decimators.  In this sense, the
	%numerical results from\cite{Crochiere1975} showed that simply using three
	%stages is a very efficient approach in practice.  Therefore, using three or at
	%most four stages for the decimators, in practice seems to cover nearly all DSP
	%applications.  

	The theory developed in this paper applies to both interpolation and decimation
	systems, but the focus will be on decimation systems, as this is the processing
	required in the analysis filter banks used as readout systems in cryogenic
	detectors.  The application of these structures to a particular hardware
	platform is also investigated by mean of examples, in order to test its
	performance in terms of number of multiplications, additions and resource
	utilization.  The design steps and some test results are shown for a Red Pitaya
	125-14 board~\cite{redpitaya}.

	\section{Digital Filter Banks}\label{FB}
	Filter banks are generally categorized as two types, \textit{analysis} and
	\textit{synthesis} filter
	banks~\cite{Bellanger1976,Vaidyanathan1990a,RonaldE.Crochiere527}. One is
	basically the inverse of the other. The
	\figurename~\ref{fig_pfb_ana_synth_bis} show the structure of a $K$-channel
	analysis and synthesis filter bank. The $x(n)$ signal is the FDM signal
	containing the channels of interest.  The combination of delays, plus
	down-sampling by $M$ ($\downarrow \negmedspace M$) delivers one input sample
	to each of the $H_k(z)$ filters at a time and at a low sampling rate ($m$).
	Here $k$ is the channel number (or the sub-band number) and $z$ is the
	$z$-transform variable.  In the analysis filter bank, every output channel
	$X_k(m)$ is recovered at the output of a M-point DFT block.  Here, the $n$ and
	$m$ (with $n >  m$) indexes denote a signal working at the input and output
	sampling rate, respectively. In the synthesis, the inverse process occurs.

	Both systems involve first the design and then the partition of a
	\emph{prototype} filter, where the partitioning occurs over a bank of smaller
	filters. An analysis FB consists of a set of filters, with system functions
	$H_k(z)$, arranged in parallel.  The frequency response characteristics of
	this filter bank splits the input signal into a corresponding number of
	sub-bands. On the other hand, a synthesis FB consists of a set of filters with
	system functions $F_k(z)$, also arranged in a parallel bank.  The output of
	these filters are summed to form the synthesized signal $\hat{x}(n)$.  The
	approach in these systems is, in general, to find the perfect reconstruction
	(PR) condition~\cite{Vetterli1987,RonaldE.Crochiere527}, where $x(n) =
	\hat{x}(n)$, with no more distortion other than a shift in time and some scale
	of amplitude.

	One of the most important classes of filter banks is the DFT filter bank
	(DFT-FB), which belongs to a more general class called modulated filter
	banks~\cite{NorbertFliege633}.  The main advantage of using modulated FB lies
	in the fact that they can be implemented using polyphase filters and an
	efficient signal processing algorithm such as the fast Fourier transform
	(FFT), resulting in significant reduction of computational
	complexity~\cite{RichardG.Lyons85}.  However, the DFT-FB can be considered a
	special case of the generalized DFT filter bank (GDFT-FB) in which the central
	frequencies and phases of the sub-bands can be more explicitly controlled. The
	GDFT-FB is based in the generalized DFT transform pair
	\begin{subequations}\label{eq_def_gdft}
		\begin{align}
			Y^{\text{(GDFT)}}_k &=\sum_{n=0}^{K-1}y(n)W_K^{(k+k_0)(n+n_0)},\label{gdft_fft}\\
			\intertext{and}
			y(n) &=\frac{1}{K}\sum_{k=0}^{K-1}Y^{\text{(GDFT)}}_k W_K^{-(k+k_0)(n+n_0)},\label{gdft_ifft}
		\end{align}
	\end{subequations}
	where $n_0$ is the reference for the time origin and $k_0$ corresponds to the
	reference for the discrete frequency origin.  The parameter $n_0$ will be equal
	to 0 in this paper, as the time reference is not modified. The parameter $W_K$ 
	is known as the \emph{twiddle factor}, and is defined as 
	\begin{equation}\label{eq_twiddle_factor}
		W_K=e^{-j2\pi/K}=\sqrt[K]{1}.
	\end{equation}
	The generalized DFT filter bank offers flexible configurations for phase
	shifting and channel stacking, and in some applications the use of a more
	flexible GDFT-FB system is preferred over a
	DFT-FB~\cite{Weib1998b,Weib1998c,Peng2013}.
	\begin{figure}[!ht] 
		\centering
		\includegraphics[width=0.7\textwidth]{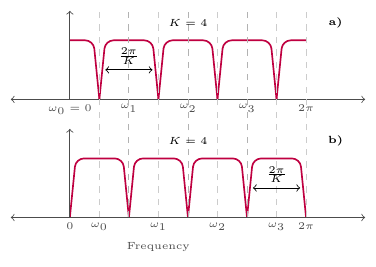}
		\caption{Partitioning of the frequency spectrum in a uniform ($K=M$)
		filter bank, where there is no overlapping between adjacent
		channels~\cite{RonaldE.Crochiere527}. The frequencies, $\omega_k$, are
		obtained by uniformly shifting the frequency response of the prototype
		filter in integer multiples of $2\pi/K$, where $K=4$ is the total number
		of channels. $M$ is the downsampling ratio.  Two different ways of
		channel stacking are shown: (a) even stacking ($k_0=0$) and (b) odd
		stacking ($k_0=1/2$).}
		%The case of a uniform ($K=M$) filter bank is shown, where there is no
		%overlapping between adjacent channels~\cite{RonaldE.Crochiere527} and
		%the number of channels is$K=4$.}
		%\caption{Different ways of filter stacking for modulated filter banks:
		%(a) even stacking ($k_0=0$) and (b) odd stacking ($k_0=1/2$). The case
		%of a uniform ($K=M$) filter bank is shown, where there is no overlapping
		%between adjacent channels~\cite{RonaldE.Crochiere527} and the number of
		%channels is$K=4$.}
		\label{fig_odd_even_stacking}
	\end{figure}

	Modifying the frequency origin, $k_0$, leads to designs with different channel
	stacking arrangements. For example, choosing the frequency origin $\omega =
	\pi/K$ (or equivalently $k_0=1/2$), for a filter bank of $K$ channels produces an
	\emph{odd} channel stacking arrangement (see
	\figurename~\ref{fig_odd_even_stacking}), such that the center frequencies of
	the channels are located at the frequencies
	\begin{equation}\label{centro_banda_impar_rep}
		\omega_k=\frac{2\pi k}{K}+\frac{\pi}{K}=\frac{2\pi
		}{K}\left(k+\frac{1}{2}\right),\quad k= 0,1,\ldots,K-1.
	\end{equation}

	An analysis filter bank consisting of $K$ filters $h_k(n),\quad k=0, 1,...,
	K-1$ is called a \emph{critically sampled} or \emph{uniform} GDFT filter bank if
	$h_k(n)$ are derived from a prototype filter $h(n)$, where
	\begin{equation}\label{eq_pf_filter}
		h_k(n) = h(n)W_K^{-(k+k_0)n},\quad k=1,2,...,K-1.
	\end{equation}
	Hence, the frequency response characteristics of the filters $h_k(n)$ are simply
	obtained by uniformly shifting the frequency response of the prototype filter in
	integer multiples of $2\pi/K$~\cite{JohnG.Proakis663}.  In the uniform FB, $K=M$
	or, equivalently $I=K/M=1$, where $M$ is the downsampling ratio of the filter
	bank and $I$ is defined as the \emph{oversampling ratio}. The frequency spectrum
	is then partitioned in a uniform manner, as illustrated in 
	\figurename~\ref{fig_odd_even_stacking}. The sub-band width $\Delta_k =
	\frac{2\pi}{K}=\frac{F_s}{K}$ is identical for each sub-band and the band
	centers are uniformly spaced at intervals of $\frac{2\pi}{K}=\frac{F_s}{K}$,
	with a system working at a sampling rate $F_s$. 

	In general, the prototype filter can be either a FIR or IIR filter. The
	selection between them depends on the application's requirements. One of these
	requirements, common in radio-astronomy and communications, where the signal is
	supposed to pass without distortion, is the linear phase of the digital
	processing chain and is one of the key points for choosing the FIR approach for
	the architectures in this paper.   

\begin{figure}[!ht]
		\centering
		\includegraphics[width=0.7\textwidth]{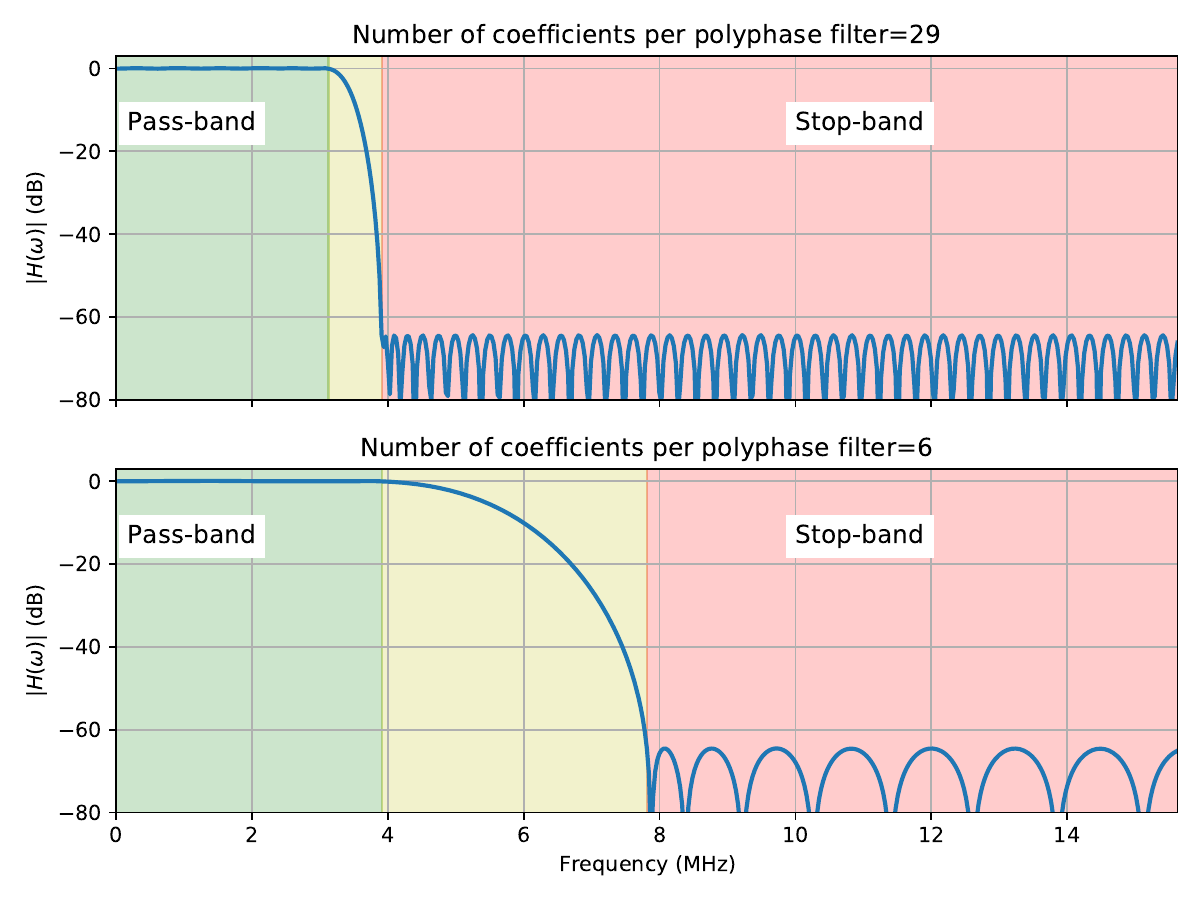}
		\caption{
			Frequency response comparison between the prototype filter designed
			for the critically sampled (top) and oversampled (bottom) cases. The
			relaxation of the specifications in the oversampled case allows for
			obtaining computational and hardware savings in the FPGA.
			}
			\label{fig_proto_compa}
	\end{figure}

	While the critically sampled GDFT-FB architecture is straightforward and
	reliable, the low-pass prototype specifications require that the stop-band and
	the transition band do not exceed the Nyquist frequency of the decimated
	sub-band, if the effect of aliases is to be minimized. If the sub-band filter
	exceeds the Nyquist frequency, aliasing can become a problem by introducing
	correlated noise that distorts the signal.  To avoid the problem of aliasing in a
	critically sampled design, it is necessary to specify that less of the nominal
	sub-band bandwidth be available for the signal, thus creating a wider guard band
	between signals on adjacent channels. This can be seen in
	\figurename~\ref{fig_proto_compa} (top), where the guard band between channels
	represents 25\% of the useful bandwidth for the channel signals. It may also be
	necessary to design a high-order prototype filter so that a sharper filter
	transition can be achieved, trying to minimize the overlap between a filter and
	its images. These solutions have the disadvantage of reducing the useful
	bandwidth of the signal or increasing the use of resources due to the higher
	number of filter coefficients and a higher computational load.  This type of
	design, although it is optimal in the sense of resource usage, may not be the
	best solution for reading superconducting detectors, precisely because there are
	guard bands between channels that make the reading system lose part of the
	spectrum. This is a problem since, for instance an MKID detector by construction
	may not resonate at the design frequency~\cite{McHugh2012,VanRantwijk2016} and
	its resonance could fall into some guard band, thus losing the corresponding
	pixel in the readout.

	To avoid reducing the available signal bandwidth or increasing unnecessarily the
	order of the prototype filter, an oversampled filter bank (GDFT-FB-OVS) can be
	used. In an oversampled design, the Nyquist frequency of the sub-band is now
	greater than the sub-band spacing ($F_s/K$). Therefore it is
	possible to have sub-band filters that overlap in terms of the input signal but
	whose images do not overlap and therefore do not cause aliasing after
	decimation~\cite{RonaldE.Crochiere527}.

	\begin{figure*}[!ht]
		\centering
		\includegraphics[width=0.434\textwidth]{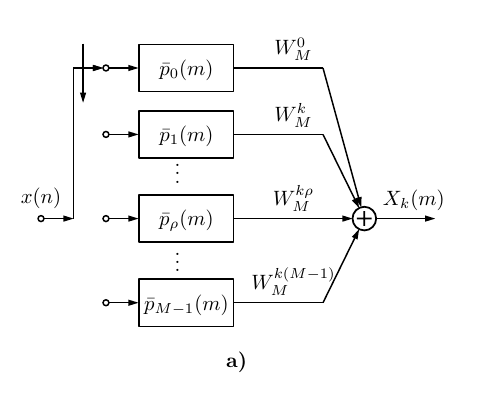}
		\includegraphics[width=0.49\textwidth]{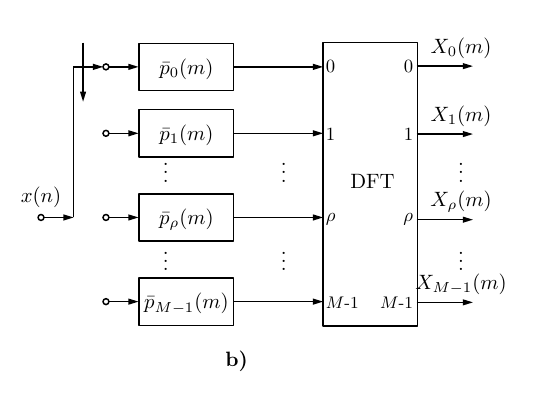}
		\caption{
			\textbf{a)} Polyphase structure of the $k$-th channel in a uniform GDFT-FB, where
			the $\bar{p}_\rho(m)$ notation was used (in place of the more generic
			$\bar{p}_{\rho,k}(m)$, where the channel number $k$ is explicitly shown),
			and \textbf{b)} structure of the polyphase filter bank, with a DFT modulator
			implemented through the FFT algorithm. Both diagrams shows a clockwise
			commutator~\cite{NorbertFliege633,Vaidyanathan1990a,RonaldE.Crochiere527}
			implementation of the uniform GDFT-FB system.
			}
			\label{fig_7_11}
	\end{figure*}
	\subsection{Polyphase structure of GDFT filter banks}\label{sec:polyphase_structure_of_DFT_PFB}
	In the implementation of digital FB, computational efficiency can be achieved by
	the use of polyphase filters~\cite{Bellanger1974,Bellanger1976} for decimation
	and interpolation.  For an analyzer, uniform GDFT-FB with $k_0=0$, it can be
	defined a set of $K$ polyphase filters $\bar{p}_\rho(m)$~\footnote{Using the
	\emph{type-3} polyphase representation. For details, see~\cite{NorbertFliege633}
	and \cite{RonaldE.Crochiere527}.}, corresponding to the
	responses to the unit impulse of the bandpass filters $h_k(n)$ for the polyphase
	branches, like
	\begin{equation}\label{p_ro}
		\bar{p}_\rho(m)=h_k(mM-\rho),\quad\rho=0,1,\ldots,M-1
	\end{equation}
	and the corresponding set of down-sampled input sequences for each channel $k$ 
	\begin{equation}\label{x_ro}
		x_\rho(m)=x_k(mM+\rho),\quad\rho=0,1,\ldots,M-1
	\end{equation}
	where $\rho$ and $k$ refers to the $\rho$-th branch of the polyphase structure
	and to the $k$-th channel of the filter bank, respectively (see
	\figurename~\ref{fig_7_11}). Here $m$ represents the new sampling rate and $n >
	m$ in the case of decimation.

\begin{figure}[!ht]
		\centering
		\includegraphics[width=0.7\textwidth,scale=1.5]{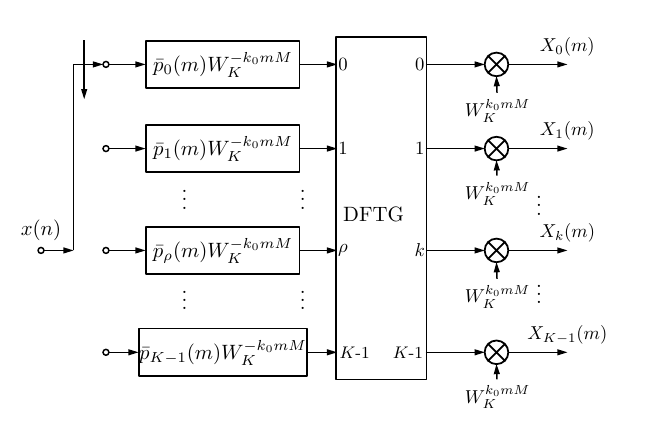}
		\caption{
			Analysis polyphase structure in an oversampled GDFT filter bank
			(assuming $n_0=0$, see text).
			}
			\label{fig_gdft_poly}
	\end{figure}

	Alternatively, for the more general case of $K=MI$ and $k_0\neq 0$, it can be
	defined a set of $K$ polyphase filters as
	\begin{equation}\label{filtro_poli_p_gdft_sm}
		\bar{p}_\rho^{(\text{GDFT-OVS})}(m)=\bar{p}_\rho(m)W_K^{k_0(\rho-mM)}
	\end{equation}
	where the $\bar{p}_\rho(m)$ factors are those defined by Eq.~\eqref{p_ro}. Using the
	polyphase filters of Eq.~\eqref{filtro_poli_p_gdft_sm} leads to the structure shown in
	\figurename~\ref{fig_gdft_poly} for a GDFT-FB-OVS analyzer. In this structure,
	the channel signal can be written as (see the Appendix for a derivation)
	\begin{equation}\label{senial_canal_total_pb_gdft_K_dif_M_fir}
		\begin{aligned}
			X_k(m) &=
			W_K^{k_0mM}\sum_{\rho=0}^{K-1}W_K^{k\rho}[x_\rho(m)\**(\bar{p}_\rho(m)W_K^{k_0(\rho-mM)})],
			\\
			k &=0,1,\ldots,K-1.
		\end{aligned}
	\end{equation}
	For large values of $M$, the FFT algorithm provides an efficient mean for
	computing the DFT and makes this approach attractive for the implementation in
	programmable logic.

	\section{FPGA implementation of an oversampled system}\label{sec:pf_design}
	In this section the FPGA implementations of the filter banks analyzed so far
	are developed. The implementations are made in the Zynq~\cite{ds190} family
	of Xilinx FPGAs, using the Vivado IDE (2018.3 version) toolkit and the
	reusable intellectual property (IP) block libraries, as well as standard VHDL
	language. Two IP blocks critical in the realization of these digital
	multi-rate systems are the FIR Compiler~\cite{pg149} and the FFT~\cite{pg109}
	block. It is used the Red Pitaya 125-14 board as the target platform for the
	tests and implementation.

	%The design of a FB starts by specifying the prototype
	%filter. The focus now is on the understanding and development of the prototype
	%filter for the polyphase GDFT-FB. In an analysis FB, the sample rate for the design
	%process is the input rate, which turns out to be the highest sample rate the
	%system will ever operate at. This frequency is $F_s=125\,\text{MHz}$ for each
	%of the two input channels in the Red Pitaya 125-14 board.  In the uniform
	%analyzer, each channel will be $1/K^{th}$ of the total received bandwidth.  By
	%partitioning the filter, what is done is to split the original filter into $K=M$ 
	%pieces. Then building the prototype means building a filter capable of filtering
	%a single channel at the full bandwidth of all $K$ channels.  The range of
	%frequencies to be preserved in the prototype filter will be $\pi/K$, for both
	%types of implementation, the critical sampling and oversampling.

	A design limitation of the FIR Compiler IP block is that it requires the
	decimation factor and the number of polyphase components to match.

	The general structure of the GDFT-FB-OVS system is very similar to critically
	sampled designs, such as the one shown in \figurename~\ref{fig_7_11}. However,
	two significant changes must be applied to the filter bank. In the first place,
	in the oversampled system the decimation factor no longer coincides with the
	number of channels because an oversampling ratio $I \neq 1$ appears in the
	equations, since in this case $ I = K / M> $ 1. For this
	reason, the implementation of filtering in the filter bank must be re-designed.

	The second change is in the design of the output frequency change state machine.
	If Eq.~\eqref{senial_canal_total_pb_gdft_K_dif_M_fir} and
	\figurename~\ref{fig_gdft_poly} are inspected, and if it is taken into account
	that we have done $n_0 = 0$ to obtain
	Eq.~\eqref{senial_canal_total_pb_gdft_K_dif_M_fir}, it can be seen that the
	output of the GDFT is followed by two separate complex multiplications. The
	first one, the phase shift operation, is simplified to multiplication by 1, that
	is, it does not require operation, since $n_0=0$. The second is an operation of
	frequency change ($k_0$) required to shift the output sub-band down so that it
	is centered in $\omega = 0$. This corresponds to a mixing operation, but a close
	examination of the possible multiplier values shows that it can be implemented
	efficiently with a state machine.
	The design of the GDFT-FB-OVS requires modifications with respect of that of the
	critically sampled GDFT-FB, since the oversampled signals result in additional
	possible values in the multiplications and therefore additional states in the
	state machine.
	%\begin{figure}[!ht]
	%\centering 
	%\includegraphics[width=0.5\textwidth]{c4_prototipo} 
	%\caption{Magnitude and phase response of the prototype filter. The normalized
	%				end of the pass-band frequency is$f_{pb}=0.05625\,\text{rad/sample}$
	%				and the normalized start of the stop-band frequency is$f_{sb} =
	%				0.0625\,\text{rad/sample}$. The maximum pass-band ripple is set to
	%				$0.1\,\text{dB}$and the minimum attenuation in the stop-band is
	%				$90\,\text{dB}$.}
	%\label{fig:pf_response}
	%\end{figure}

	\subsection{Oversampled polyphase decomposition}\label{descomposicion_poli_ovs}
	In general, several implementations of oversampled filter banks with even
	channel stacking can be found in the literature~\cite{Berner1999, Satorius2006,
	DeLeon1998}. They all use a similar structure, having integer interpolators in
	the sub-bands after the operation of a commutator.  These designs use the same
	commutator as in the critically sampled design.  Thus, an integer interpolator is
	applied to each channel just after the commutator.  With this approach, it is
	possible to oversample the frequency in each channel by filling in the input
	samples with $I-1$ zeros.

	In this paper, an alternative FPGA implementation of an oversampled polyphase
	filter bank is developed, based on the theory shown in
	Section~\ref{sec:polyphase_structure_of_DFT_PFB}. This model allows for
	additional re-configurations to implement a GDFT-FB-OVS system with IP blocks
	from the standard library.

	\subsection{Filters for the oversampled system}\label{fir_sm}
	Internally, the FIR Compiler block implements a critically sampled polyphase
	decimation structure, as using a commutator~\cite{pg149}. So both the decimation
	factor and the possible commutator paths in the GDFT-FB system must be equal to the
	number of channels.  However, in a GDFT-FB-OVS system, the decimation factor
	must, by definition, be less than the number of channels, since $K/M>1$.

\begin{figure}[!ht]
		\centering
		\includegraphics[width=0.7\textwidth]{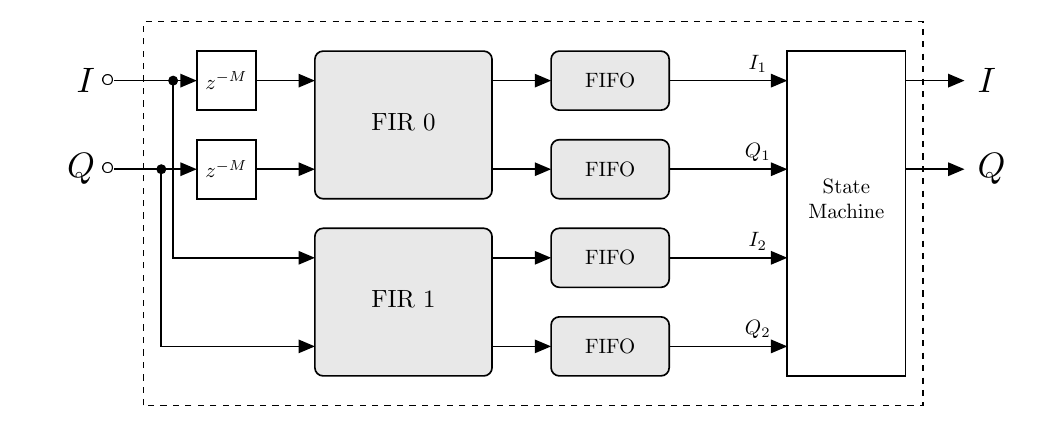}
		\caption{
			Polyphase decimation filter implemented using real or complex
			critically sampled polyphase decimating filters, depending on whether
			it is a GDFT-FB or GDFT-FB-OVS system.
			}
			\label{fig_fir_sm}
	\end{figure}

	To better understand the changes required in a GDFT-FB-OVS, it is convenient to
	express and analyze the equation that describes the polyphase representation of
	the filters in the domain of the variable $z$. If we start from the type-3
	polyphase representation of the prototype filter
	\begin{equation}\label{eq_transformada_z_polifasica_t3_c4}
		H(z) = \sum_{n=-\infty}^{\infty} h(n)z^{-n} = \sum_{\rho = 0}^{K-1}z^{\rho}\,P_\rho(z^{K})
	\end{equation}
	where
	\begin{equation}\label{eq_rep_poly_z}
		P_\rho(z) = \sum_{m=-\infty}^{\infty}\bar{p}(m)z^{-m} =
		\sum_{m=-\infty}^{\infty}h(mK-\rho)z^{-m}.
	\end{equation}
	Now, to find an alternative expression, valid for its application in the FIR
	Compiler block and taking into account an oversampling relation $I = 2$, we
	must expand Eq.~\eqref{eq_transformada_z_polifasica_t3_c4} to
	obtain Eq.~\eqref{expansion_ec_poly}. 
	%\begin{figure*}[!h]
	%\begin{widetext}
	%	\mbox{
			\begin{equation}\label{expansion_ec_poly}
				\begin{aligned}
					H(z) &= P_0(z^K) + z^1 P_1(z^K) + z^2
					P_2(z^K)+\ldots+z^{K-1}P_{K-1}(z^K) \\
					&= P_0(z^K) + z^1 P_1(z^K) +\ldots+ z^{(K/2-1)} P_{K/2-1}(z^K) \\
					&+ z^{(K/2)}\left[P_{K/2}(z^K) + z^1 P_{K/2+1}(z^K) +\ldots+
					z^{(K/2-1)} P_{K-1}(z^K)\right].
				\end{aligned}
			\end{equation}
	%		}
	%\end{widetext}
	%\end{figure*}
	This can be generalized, for any integer $I$ and taking into account that $M =
	K/I$, as 
	\begin{equation}\label{expansion_ec_poly_general}
		H(z) = \sum_{r=0}^{I-1}z^{rM}\left[\sum_{\rho=0}^{M-1}z^\rho
		P_{\rho+rM}(z^K)\right] = \sum_{r=0}^{I-1} z^{rM} P_{FIR_r}(z)
	\end{equation}
	where
	\begin{equation}\label{expansion_ec_poly_fir}
		P_{FIR_r}(z) = \sum_{\rho=0}^{M-1}z^\rho P_{\rho+rM}(z^K), \quad
		r=0,1,\ldots,I-1.
	\end{equation}
	With this decomposition, the channels are divided into $I$ groups, defined by
	$P_{FIR_r}(z)$, and there are $M$ channels in each group
	(\figurename~\ref{fig_fir_sm}).  The benefit of this grouping is that the number
	of channels in each group is now equal to the downsampling factor $M$.
	Therefore, in each group, a uniformly sampled decomposition is processed, making
	it compatible with the FIR Compiler blocks.

	Analyzing the problem from another perspective, it can be seen that the whole
	system achieves a x2 oversampling, because two commutators are taking the data at
	the input sampling frequency in parallel, therefore the whole system is getting
	twice the number of input samples simultaneously.

	So the general solution is that oversampled polyphase decimation is implemented
	using $I$ polyphase decimation FIR blocks, which can be real or complex as
	needed for the DFT-FB or GDFT-FB system, respectively. From the original
	prototype filter, the subgroups of interpolated polyphase filters are created
	according to Eq.~\eqref{expansion_ec_poly_fir}.  In the FPGA
	implementation, each FIR Compiler block is loaded with one of these groups of
	interpolated coefficients from the original prototype filter.

	In general, most applications only require an oversampling ratio of $ I = 2 $,
	because oversampling by 2 can already greatly reduce aliasing with adjacent
	channels. Raising the sub-sampling factor to numbers greater than 2 will require
	more FIR Compiler IP blocks with a consequent increase in calculations and
	hardware usage.

	\subsection{State machine and reordering of the samples}\label{sm_reorder}
	In the oversampled system, it is necessary to add a FIR block selector state
	machine since each of the $I$ FIR blocks executes and produces its outputs in
	parallel. This state machine will be in charge of multiplexing the outputs of
	the individual FIR Compiler blocks in time to be able to feed the correct data
	(real and imaginary part) to the FFT IP block. Specifically, to implement the
	DFT-FB or GDFT-FB system correctly, the outputs of the oversampled FIRs from
	branch 0 to branch $K$ of the general polyphase decomposition must be supplied
	sequentially to the FFT block. First the $M$ samples 0 to $M-1$ are selected
	from the FIR$_1$, then the next $M$ samples, from $M$ to $2M-1$ are selected
	from the FIR$_2$, and so on, until the final $M$ samples, from $(I-1)M$
	to $IM-1$ are selected from the FIR$_I$, at which point the sequence begins
	again.

	The \figurename~\ref{fig_fir_sm} shows the block diagram implementation of
	the oversampled polyphase decimation FIR using 2 FIR Compiler blocks, FIFO
	memory blocks for buffers and a selector state machine. For the case $I = 2$,
	this selector state machine becomes a simple multiplexer with two inputs and one
	output.

	As in the case of the critically sampled GDFT-FB system, the outputs of the
	sub-bands of the IP FFT block require a frequency shift to re-center each output
	channel at $\omega = 0$. As before, this final frequency change can be
	implemented using a state machine, albeit with more states. The number of states
	depends on the oversampling factor $I$.
	Substituting $k_0 = 1/2$ and taking into account that $K = MI$
	in Eq.~\eqref{senial_canal_total_pb_gdft_K_dif_M_fir} we obtain
	\begin{equation}\label{fact_multiplicativo_sm}
		W_K^{k_0mM} = e^{-j\frac{2\pi}{K}(k_0mM)} = e^{-j\frac{\pi}{I} m},\quad m \in
		\mathbb{N},
	\end{equation}
	which in the case of $I = 2$ reduces to
	\begin{equation}\label{fact_multiplicativo_sm_opt}
		W_K^{k_0mM} = \left\{ \begin{array}{rcl}
			1 & & m = 4r,\\
			-j & & m = 4r+1,\\
			-1 & & m = 4r+2,\\
			+j & & m = 4r+3, \quad r,m \in \mathbb{N}
		\end{array}\right.
	\end{equation}
	that is, the required frequency change operation can be replaced by a state
	machine with 4 states. These multiplications can be implemented without any
	multiplier, since the operations are equivalent to pass unchanged, negate, or
	interchange the in-phase and quadrature components of the data. While similar
	state machines can be derived for larger oversampling factors, they will require
	some multipliers in this case. This is perhaps another reason to consider
	avoiding higher oversampling ratios.

	%The procedure of how the state machine works is shown in
	%\figurename~\ref{fig_maquina_estados_K_dif_M}. All processing in the state
	%machine is done on
	%the in-phase and quadrature components of the channels. The count of samples
	%entered into the state machine is carried out by means of a dedicated counter.
	%The operation of the machine is regulated by the line \verb!T_valid! of the IP
	%FFT block. This line indicates that there is valid data at the output of the FFT
	%block and then the counter is enabled and the state machine begins to receive
	%data. If the state machine is running in the first round of $ K $ samples,
	%$X(K)\text{a} X(K-1)$, multiply these samples by 1. If the state machine is
	%receiving the set of samples from $ X (K) \text{a} X (2K-1) $, it will
	%multiply the samples by -j. If the state machine is receiving the set of samples
	%from $ X(2K)\text{a} X(3K-1)$, it will multiply these samples by -1. If
	%the state machine is receiving the set of samples from $ X (3K)\text{a}
	%X(4K-1)$, it will multiply these samples by j. After the machine has processed
	%$4K$ samples, the counter resets.
	%\begin{figure}[!ht]
	%\centering
	%        \includegraphics[width=0.49\textwidth]{c4_maquina_estados_K_dif_M_mio}
	%        \caption{Flujo de trabajo de la máquina de estado de cambio de
	%        frecuencia para el sistema sobre-muestreado.}
	%\label{fig_maquina_estados_K_dif_M}
	%\end{figure}

	\section{FPGA verification}\label{fpga_verification}
	The design start with the selection of the type of filter we want for the
	application, which for this design will be an equi-ripple low-pass FIR
	filter~\cite{Parks1972}. Then we set the sample rate ($F_s$), the end of the
	pass-band frequency ($F_{pb}$), and the start of the stop-band frequency
	($F_{sb}$), along with the desired stop-band attenuation ($\delta_s$) and the
	maximum pass-band ripple ($\delta_p$) in dB.  These design parameters can be
	expressed also in normalized frequencies (lowercase letters), where $F_s/2 =
	62.5\,\text{MHz}\equiv 1\,\text{rad/sample}$, $f_{pb} =
	0.9\times(1/16)\,\text{rad/sample}$ and $f_{sb}=1/16\,\text{rad/sample}$, for
	example in a critically sampling design.  Table~\ref{table_proto_parameters}
	shows the design specifications for the uniform and oversampled prototype
	filters. 
	\begin{table}[!t]
		\caption{Design specifications for the critically sampled and oversampled
		systems.}
		\label{table_proto_parameters}
		\centering
		\begin{tabular}{|l|l|l|}% p{2.2cm}| p{2.2cm}|}
			\hline
			&\textbf{GDFT-FB} &\textbf{GDFT-FB-OVS}\\
			\hline
			\textbf{Parameter} &\textbf{Value} &\textbf{Value}\\
			\hline\hline
			Type                          & FIR & FIR\\
			Style                         & Low-pass & Low-pass\\
			Sampling rate ($F_s$)         & 125\,MHz & 125\,MHz\\
			Pass-band ($F_{pb}$)          & 3.12500\,MHz & 3.90625\,MHz\\
			Stop-band ($F_{sb}$)          & 3.90625\,MHz & 7.81250\,MHz\\
			Pass-band ripple ($\delta_p$) & 0.1\,dB & 0.1\,dB\\
			Stop-band atten. ($\delta_s$) & 66\,dB & 66\,dB\\
			\hline\hline
			Channel number (K)      &16&16\\
			Downsampling factor (M) &16&8\\
			Oversampling ratio (I)  &1&2\\
			\hline
		\end{tabular}
	\end{table}

\begin{figure*}[!ht]
		\centering
		\includegraphics[width=0.8\textwidth]{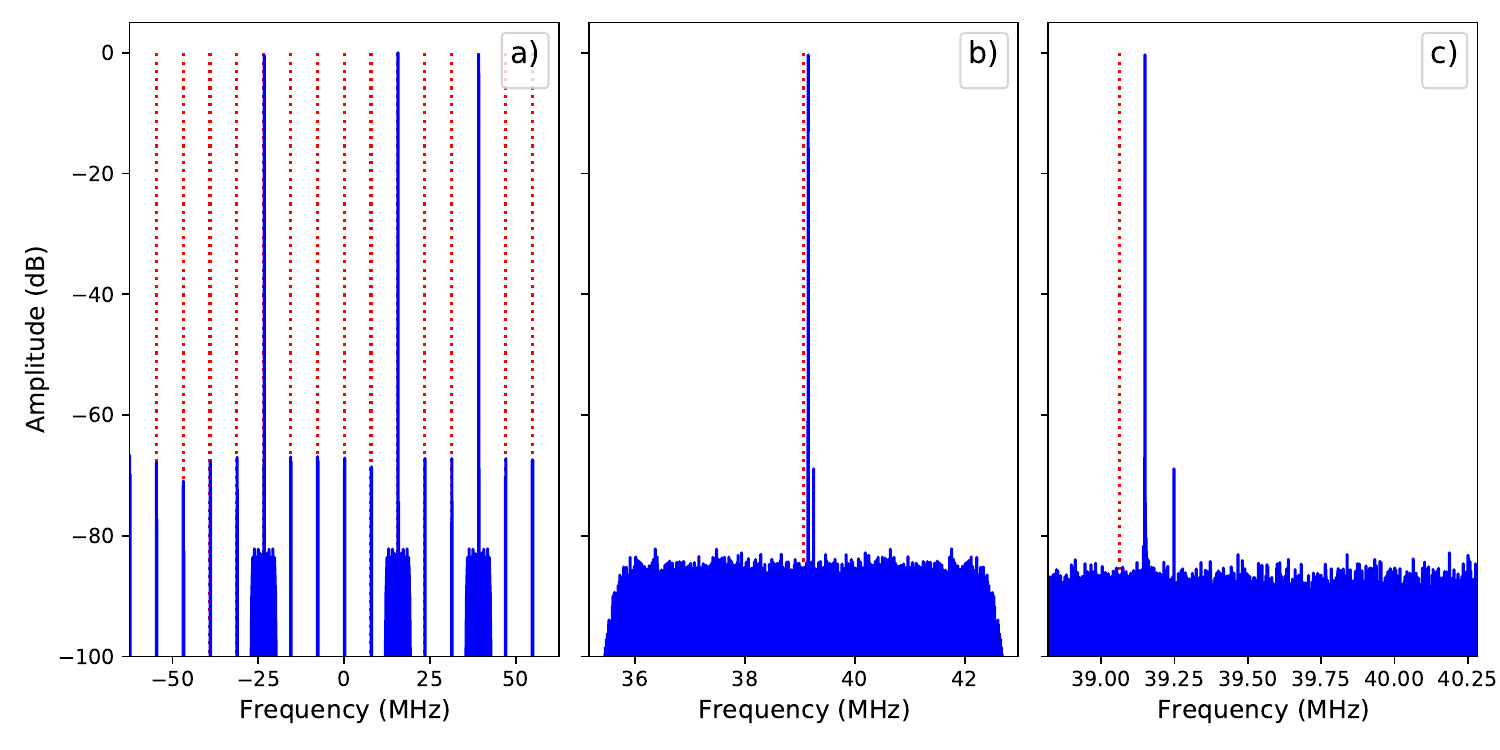}
		\caption{\textbf{a)} Normalized input spectrum for the 16-channel GDFT-FB
		systems. The dotted lines show the channel centers.  There are three input
		tones at channel 3, 6 and 14, respectively. A close view of the channel 6 is
		shown in \textbf{b)} and \textbf{c)}, where the tone at frequency
		$f=1.385250\,\text{MHz}$ is shown after the mixing up in the 16-channel
		synthesizer. 
			}
			\label{zoom_in_spectrum}
	\end{figure*}

\begin{figure*}[!ht]
		\centering
		\includegraphics[width=0.8\textwidth]{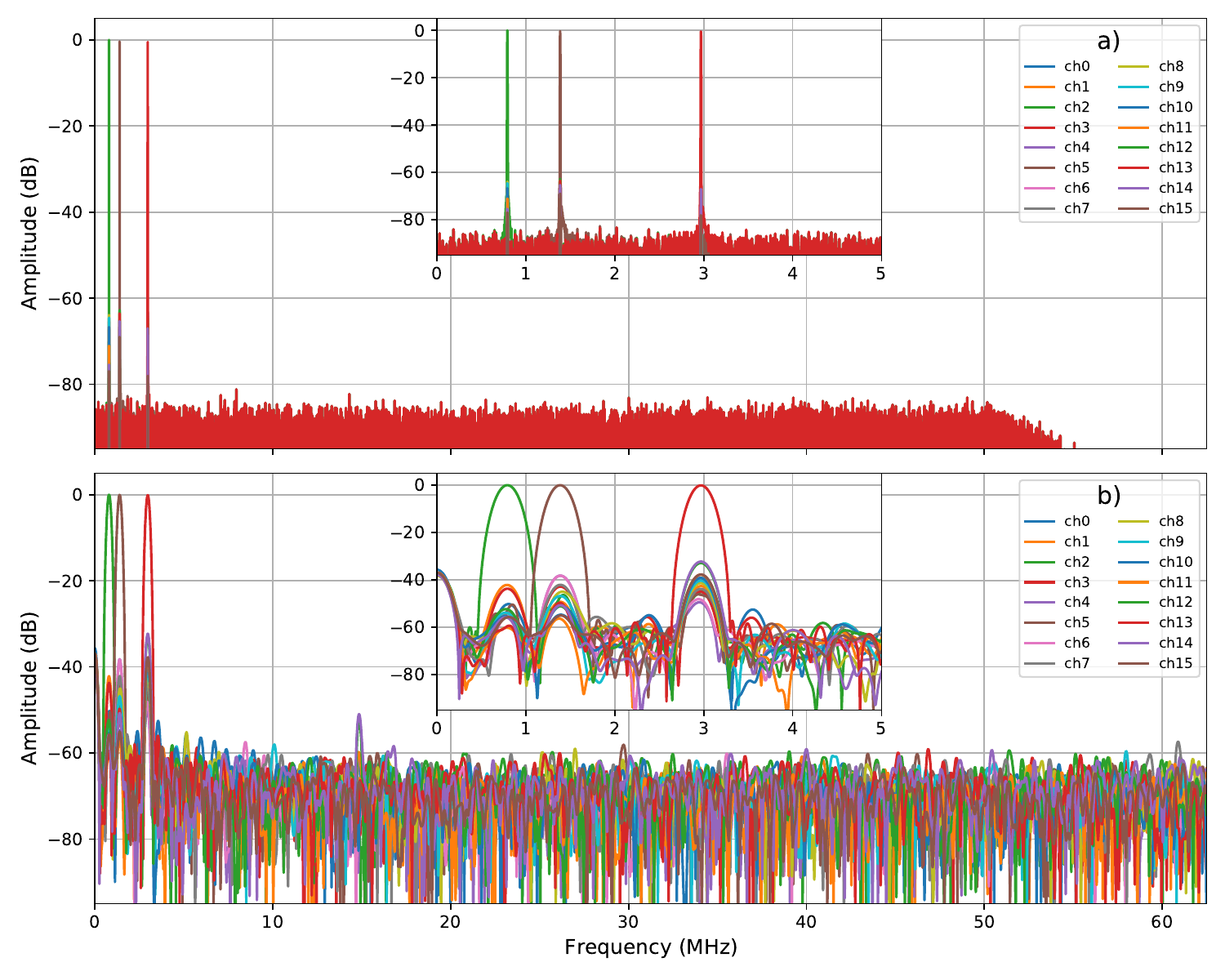}
		\caption{
			\textbf{a)} Simulation of the normalized output spectrum for the critically sampled
			16-channel GDFT-FB analyzer. A noise margin of $\sim 60\,\text{dB}$ is
			observed between adjacent channels. \textbf{b)} Normalized frequency
			response of the same system, applied to the Red Pitaya board. The
			noise margin decreases to $\sim 35\,\text{dB}$, due to external noise
			added by the electronics on the board. The insets shows a close
			view of the frequency range 0-5\,MHz, where it is shown the three
			recovered tones after passing through the 16-channel analyzer. The tones
			at frequecies $f_3=0.791571\,\text{MHz}$, $f_6=1.385250\,\text{MHz}$ and
			$f_{14}=2.968394\,\text{MHz}$ pass through the 16-channel analizer in
			channels 3, 6 and 14, repectively.
			}
			\label{fig_fdm_out_rxc16_dft}
	\end{figure*}

	The design was synthesized and implemented in the Zynq FPGA of the Red Pitaya
	board. The input spectrum is shown in \figurename~\ref{zoom_in_spectrum},
	where a close view of the channel 6 is also shown. The output of the testings
	can be seen in \figurename~\ref{fig_fdm_out_rxc16_dft} and
	Table~\ref{table_uso_hw_dftg} for the 16-channels GDFT-FB and GDFT-FB-OVS systems.
	\figurename~\ref{fig_fdm_out_rxc16_dft} shows a comparison of the simulated
	signals and the acquired signals for a GDFT-FB system. 
	%These last two figures show the recovering
	%of the input tones in the time domain, also showing the proper functioning of
	%the design.
	%ver el paper de rondeau_gr_filterbank ahi explica un poco lo que hacen el
	%filtro de analisis y el de sintesis

	%El script para hacer esta grafica está en
	%/home/arnaldi/work/gitrepos/lharnaldi/channelizer/dft_rxc16/python
	\subsection{Hardware resources utilization}\label{uso_hardware}
	The wideband input signal was digitized with a resolution of 14 bits per sample.
	The coefficients of the FIR Compiler block, most of the input and output
	memories, and the intermediate results of the FFT will generally require the use
	of blocks of RAM. All multiplications and additions were implemented in DSP48
	blocks for maximum performance.

	Therefore, the use of the RAM and DSP48 memory blocks are the most important to
	evaluate. Table~\ref{table_uso_hw_dftg} shows the FPGA resource usage for
	the 16-channel GDFT-FB system, with odd and even channel stacking and for the
	oversampled system, with odd channel stacking. It should be noted that
	Table~\ref{table_uso_hw_dftg} only compares the use of the DSP48 and BRAM blocks
	in the FPGA, without taking into account the additional hardware that some of the
	implementations may require. For example, there are selectors and FIFO memories
	associated with the GDFT-FB-OVS architecture (\figurename~\ref{fig_fir_sm}),
	which are implemented with DRAM (Distributed RAM) memories and are not analyzed
	in this table.
	\begin{table*}[!h]
		\caption{
			Use of hardware resources in 16-channel GDFT-FB and GDFT-FB-OVS systems.
			\\The percentage shown refers to the total resources available on the Red
			Pitaya 125-14 board.
			}
			\label{table_uso_hw_dftg}
			\centering
			\rowcolors{1}{}{gray!20}
			\begin{tabular}{|l||c|c|c|c|}   % c, not p{2.6cm}: p columns with xcolor[table] break revtex (Extra \or)
				\hline
				%\multicolumn{4}{|c|}{Country List} \\
				\rowcolor{LightBlue3} \cellcolor[gray]{0.80}\textbf{Parameter} &
				\multicolumn{2}{|c|}{\textbf{GDFT-FB}}&\multicolumn{2}{|c|}{\textbf{GDFT-FB-OVS}} \\
				\hline
				\rowcolor{LightBlue3} \cellcolor[gray]{0.40}&\cellcolor[gray]{0.9} $k_0=0$ &
				\cellcolor[gray]{0.8} $k_0=\frac{1}{2}$ & \cellcolor[gray]{0.9} $k_0=0$ &\cellcolor[gray]{0.8} $k_0=\frac{1}{2}$ \\
				\hline
				\rowcolor{LightBlue1} \cellcolor[gray]{0.40}&\multicolumn{4}{|c|}{FIR} \\
				\hline
				DSP Slices & 60 (75\%)& 2 $\times$ 60 (150\%) & 2 $\times$ 26 (65\%)& 2 $\times$ 26 (65\%)\\
				BRAM         & 0 (0\%)  & 0 (0\%)              & 0 (0\%)& 0 (0\%)\\
				Architecture & \shortstack{Systolic\\Multiply\\Accumulate}& \shortstack{Systolic\\Multiply\\Accumulate} & \shortstack{Systolic\\Multiply\\Accumulate}& \shortstack{Systolic\\Multiply\\Accumulate}\\
				\hline
				\rowcolor{LightBlue1} \cellcolor[gray]{0.40}&\multicolumn{4}{|c|}{FFT} \\
				\hline
				DSP Slices & 12 (15\%)& 12 (15\%)& 12 (15\%)& 12 (15\%)\\
				BRAM         & 2 (3.3\%)& 2 (3.3\%)& 2 (3.3\%)& 2 (3.3\%)\\
				Implementation& \shortstack{Pipelined\\Streaming I/O}&\shortstack{Pipelined\\Streaming I/O}&\shortstack{Pipelined\\Streaming I/O}&\shortstack{Pipelined\\Streaming I/O}\\
				\hline
			\end{tabular}
	\end{table*}
	As a result, the GDFT-FB system with odd channel stacking will use 50\% more
	resources, but may have better use of spectrum. On the other hand, if the
	oversampled implementations are compared with the critically sampled ones, it
	can be stated that the resource utilization is 10\% lower. However, any of these
	GDFT-FB systems has a very high resource efficiency compared to the per channel
	approach, that is, compared to the case of implementing the filter without using
	polyphase decomposition.

	In the per channel approach, 16 channels require a 465 coefficient FIR filter
	and its corresponding digital down converter with different center frequencies.

	Finally, the GDFT-FB systems can be used to implement analyzers or synthesizers
	with odd and even channel stacks (with a suitable choice of $ k_0 $), because
	they involve a general design. However, the GDFT-FB system is the most efficient
	design as it does not require the use of complex FIR blocks (requiring 2 FIR
	Compiler blocks) or the output frequency shift state machine.

	From Table~\ref{table_uso_hw_dftg} it is also clear that if you want to further
	optimize these implementations, you will need a more detailed analysis of the
	structure of the FIR filters, since the FFT block is by definition an efficient
	implementation. The latter is evident in Table~\ref{table_uso_hw_dftg}, where
	all four implementations use the same amount of resources in their FFT block.
	%The detailed analysis of the structure of FIR filters and their optimization is
	%a problem that is addressed in the next chapter of this thesis.

	%\begin{figure*}[!ht]
	%\centering 
	%	\includegraphics[width=\textwidth]{compa_RT_NT}
	%				\caption{Partitioning of the frequency spectrum in a uniform DFT
	%				filter bank. The frequencies,$w_k$, are obtained by uniformly
	%				shifting the frequency response of the prototype filter in integer
	%				multiples of$2\pi/M$, where$M$is the total number of channels.} 
	%				%$F_s$is the sampling frequency,$N$is the number of channels and
	%				%$F_{BW}$is the bandwidth of the polyphase filters in the bank.$F_s/N$
	%				%is the separation between channels. The plot shows how the spectrum is
	%				%partitioned for the speciffic case of$N=16$and for the RedPitaya
	%				%board, where$F_s=125\,\text{MHz}$.}
	%\label{fig_spectrum_partitioning}
	%\end{figure*}
	%tomado de 073695621.pdf leo_mails

	\section{Conclusions}\label{sec:conclusions}
	This paper presented the design and implementation of a 16-channel polyphase
	filter bank (GDFT-FB and GDFT-FB-OVS), a resource-efficient algorithm for
	multirate digital signal processing.  It was shown the versatility of these kind
	of systems. Showing also, its feasibility for implementing readout architectures
	for large cryogenic detector arrays. One of the main objectives in these systems
	is the low resource usage, and the architectures shown here have demonstrated
	its efficiency in this field, with respect to the direct implementation of the
	involved algorithms in the signal processing chain.  

	It has been shown, through simulations and after implementation into an FPGA,
	its low-hardware resource usage and its capability of processing complex
	signals. The implementation was created aiming to provide the criteria to take into account and
	the design guidelines when implementing this kind of hardware in a
	reconfigurable platform. 

	The implementation of the system showed to be highly configurable and flexible,
	thanks to the use of the FPGA present in the Red Pitaya board.  The design was
	targeted to a specific platform and frequencies, but can be easily adapted to
	another range of frequencies and hardware platform. 

	%
	%\section*{Acknowledgment}
	%
	%The authors would like to thank X. Bertou and H. Asorey for their support  and
	%discussion during the preparation of this document.
	\section*{Data availability}\label{data_availability}
	The data that support the findings of this study are available from the
	corresponding author upon reasonable request.

	\appendix\label{sec:apx}
	\section{Derivation of Eq.~\eqref{senial_canal_total_pb_gdft_K_dif_M_fir}}\label{appendixa}
	In order to derive Eq.~\eqref{senial_canal_total_pb_gdft_K_dif_M_fir}, we can start
	by using Eq.~\eqref{p_ro}, where the clockwise commutator
	model~\cite{RonaldE.Crochiere527,NorbertFliege633} has been used for the
	analyzer. According to this definition, it can be shown that the branch signals
	for the analyzer are that of Eq.~\eqref{x_ro}. 

	The generic equation defining a GDFT-FB analyzer is
	\begin{equation}\label{senial_canal_total_pb_gdft}
		\begin{aligned}
			X_k(m) &= \sum_{n=-\infty}^\infty
			h(mM-n)x(n)W_K^{(k+k_0)(n+n_0)},\\
			k&=0,1,\ldots,K-1
		\end{aligned}
	\end{equation}
	where $K$ is the number of channels, $M$ is the decimation ratio of the channel
	signals $X_k(m)$, and $h(n)$ is the analysis filter. The center frequencies of
	the channels in this filter bank are at the frequencies
	\begin{equation}\label{centro_banda_gdft}
		\omega_k = \frac{2\pi}{K}\left(k+k_0\right), \quad k= 0,1,\ldots,K-1.
	\end{equation}
	Now, if Eq.~\eqref{p_ro} and Eq.~\eqref{x_ro} are applied to
	Eq.~\eqref{senial_canal_total_pb_gdft}, for the case $K = M$, $n_0=0$ and making the change of
	variables $n = rM + \rho$, the input signals of the branch are obtained as
	\begin{equation}\label{senial_canal_total_pb_gdft_K_sim_M}
		\begin{aligned}
			X_k(m)&=
			W_M^{k_0mM}\sum_{\rho=0}^{M-1}W_M^{(k+k_0)\rho}[x_\rho(m)\**(\bar{p}_\rho(m)W_M^{-k_0mM})],
			\\
			k &=0,1,\ldots,M-1,
		\end{aligned}
	\end{equation}
	where the symbol $*$ denotes convolution.

	The Eq.~\eqref{senial_canal_total_pb_gdft_K_sim_M} can be written in a more
	convenient form for the application to the FIR Compiler
	\begin{equation}\label{senial_canal_total_pb_gdft_K_sim_M_fir}
		\begin{aligned}
			X_k(m) &=
			W_M^{k_0mM}\sum_{\rho=0}^{M-1}W_M^{k\rho}[x_\rho(m)\**(\bar{p}_\rho(m)W_M^{k_0(\rho-mM)})],
			\\
			k &=0,1,\ldots,M-1.
		\end{aligned}
	\end{equation}
	That means, the polyphase filters can now be written as 
	\begin{equation}\label{filtro_poli_p_gdft}
		\bar{p}_\rho^{(\text{GDFT})}(m) = \bar{p}_\rho(m)W_M^{k_0(\rho-mM)}
	\end{equation}
	and
	\begin{equation}\label{filtro_poli_q_gdft}
		q_\rho^{(\text{GDFT})}(m) = q_\rho(m)W_M^{-k_0(\rho+mM)}
	\end{equation}
	for the analysis and synthesis, respectively.

	Finally, by setting $K = MI$ and carefully following the math, a more general
	form of the structure of the GDFT polyphase filter bank is derived, the
	oversampled GDFT filter bank.

	If we start again from Eq.~\eqref{p_ro} and use the change of variables $n = rK +
	\rho$, we can find the equation that describes the input signal to channel $k$
	in the oversampled analyzer as
	\begin{equation}\label{senial_canal_total_pb_gdft_K_dif_M}
		\begin{aligned}
			X_k(m) &=
			W_K^{k_0mM}\sum_{\rho=0}^{K-1}W_K^{(k+k_0)\rho}[x_\rho(m)*(\bar{p}_\rho(m)W_K^{-k_0mM})],\\
			k &= 0,1,\ldots,K-1
		\end{aligned}
	\end{equation}
	Here also we can find a more convenient expression to be applied to the FIR
	Compiler filters in the FPGA. If we
	reorder Eq.~\eqref{senial_canal_total_pb_gdft_K_dif_M}, we have
	\begin{equation}\label{senial_canal_total_pb_gdft_K_dif_M_fir_append}
		\begin{aligned}
			X_k(m) &=
			W_K^{k_0mM}\sum_{\rho=0}^{K-1}W_K^{k\rho}[x_\rho(m)\**(\bar{p}_\rho(m)W_K^{k_0(\rho-mM)})],
			\\
			k &=0,1,\ldots,K-1,
		\end{aligned}
	\end{equation}
	which is the same as Eq.~\eqref{senial_canal_total_pb_gdft_K_dif_M_fir}.
	That means, in this case, the polyphase filters are
	\begin{equation}\label{filtro_poli_p_gdft_sm_append}
		\bar{p}_\rho^{(\text{GDFT-OVS})}(m) = \bar{p}_\rho(m)W_K^{k_0(\rho-mM)}
	\end{equation}
	and
	\begin{equation}\label{filtro_poli_q_gdft_sm_append}
		q_\rho^{(\text{GDFT-OVS})}(m) = q_\rho(m)W_K^{-k_0(\rho+mM)},
	\end{equation}
	for the oversampled analyzer and synthesizer, respectively.

	\bibliography{mybib,library,mybooks}

\begin{thebibliography}{33}%
\makeatletter
\providecommand \@ifxundefined [1]{%
 \@ifx{#1\undefined}
}%
\providecommand \@ifnum [1]{%
 \ifnum #1\expandafter \@firstoftwo
 \else \expandafter \@secondoftwo
 \fi
}%
\providecommand \@ifx [1]{%
 \ifx #1\expandafter \@firstoftwo
 \else \expandafter \@secondoftwo
 \fi
}%
\providecommand \natexlab [1]{#1}%
\providecommand \enquote  [1]{``#1''}%
\providecommand \bibnamefont  [1]{#1}%
\providecommand \bibfnamefont [1]{#1}%
\providecommand \citenamefont [1]{#1}%
\providecommand \href@noop [0]{\@secondoftwo}%
\providecommand \href [0]{\begingroup \@sanitize@url \@href}%
\providecommand \@href[1]{\@@startlink{#1}\@@href}%
\providecommand \@@href[1]{\endgroup#1\@@endlink}%
\providecommand \@sanitize@url [0]{\catcode `\\12\catcode `\$12\catcode
  `\&12\catcode `\#12\catcode `\^12\catcode `\_12\catcode `\%12\relax}%
\providecommand \@@startlink[1]{}%
\providecommand \@@endlink[0]{}%
\providecommand \url  [0]{\begingroup\@sanitize@url \@url }%
\providecommand \@url [1]{\endgroup\@href {#1}{\urlprefix }}%
\providecommand \urlprefix  [0]{URL }%
\providecommand \Eprint [0]{\href }%
\providecommand \doibase [0]{https://doi.org/}%
\providecommand \selectlanguage [0]{\@gobble}%
\providecommand \bibinfo  [0]{\@secondoftwo}%
\providecommand \bibfield  [0]{\@secondoftwo}%
\providecommand \translation [1]{[#1]}%
\providecommand \BibitemOpen [0]{}%
\providecommand \bibitemStop [0]{}%
\providecommand \bibitemNoStop [0]{.\EOS\space}%
\providecommand \EOS [0]{\spacefactor3000\relax}%
\providecommand \BibitemShut  [1]{\csname bibitem#1\endcsname}%
\let\auto@bib@innerbib\@empty
%</preamble>
\bibitem [{\citenamefont {Day}\ \emph {et~al.}(2003)\citenamefont {Day},
  \citenamefont {LeDuc}, \citenamefont {Mazin}, \citenamefont {Vayonakis},\
  and\ \citenamefont {Zmuidzinas}}]{Day2003}%
  \BibitemOpen
  \bibfield  {author} {\bibinfo {author} {\bibfnamefont {P.~K.}\ \bibnamefont
  {Day}}, \bibinfo {author} {\bibfnamefont {H.~G.}\ \bibnamefont {LeDuc}},
  \bibinfo {author} {\bibfnamefont {B.~A.}\ \bibnamefont {Mazin}}, \bibinfo
  {author} {\bibfnamefont {A.}~\bibnamefont {Vayonakis}},\ and\ \bibinfo
  {author} {\bibfnamefont {J.}~\bibnamefont {Zmuidzinas}},\ }\bibfield  {title}
  {\bibinfo {title} {{A broadband superconducting detector suitable for use in
  large arrays}},\ }\href {https://doi.org/10.1038/nature02037} {\bibfield
  {journal} {\bibinfo  {journal} {Nature}\ }\textbf {\bibinfo {volume} {425}},\
  \bibinfo {pages} {817} (\bibinfo {year} {2003})}\BibitemShut {NoStop}%
\bibitem [{\citenamefont {Mazin}\ \emph
  {et~al.}(2006{\natexlab{a}})\citenamefont {Mazin}, \citenamefont {Bumble},
  \citenamefont {Day}, \citenamefont {Eckart}, \citenamefont {Golwala},
  \citenamefont {Zmuidzinas},\ and\ \citenamefont {Harrison}}]{Mazin2006a}%
  \BibitemOpen
  \bibfield  {author} {\bibinfo {author} {\bibfnamefont {B.~A.}\ \bibnamefont
  {Mazin}}, \bibinfo {author} {\bibfnamefont {B.}~\bibnamefont {Bumble}},
  \bibinfo {author} {\bibfnamefont {P.~K.}\ \bibnamefont {Day}}, \bibinfo
  {author} {\bibfnamefont {M.~E.}\ \bibnamefont {Eckart}}, \bibinfo {author}
  {\bibfnamefont {S.}~\bibnamefont {Golwala}}, \bibinfo {author} {\bibfnamefont
  {J.}~\bibnamefont {Zmuidzinas}},\ and\ \bibinfo {author} {\bibfnamefont
  {F.~A.}\ \bibnamefont {Harrison}},\ }\bibfield  {title} {\bibinfo {title}
  {{Position sensitive x-ray spectrophotometer using microwave kinetic
  inductance detectors}},\ }\href {https://doi.org/10.1063/1.2390664}
  {\bibfield  {journal} {\bibinfo  {journal} {Appl. Phys. Lett.}\ }\textbf
  {\bibinfo {volume} {89}},\ \bibinfo {pages} {222507} (\bibinfo {year}
  {2006}{\natexlab{a}})}\BibitemShut {NoStop}%
\bibitem [{\citenamefont {Wallraff}\ \emph {et~al.}(2004)\citenamefont
  {Wallraff}, \citenamefont {Schuster}, \citenamefont {Blais}, \citenamefont
  {Frunzio}, \citenamefont {Huang}, \citenamefont {Majer}, \citenamefont
  {Kumar}, \citenamefont {Girvin},\ and\ \citenamefont
  {Schoelkopf}}]{AndreasWallraff2004}%
  \BibitemOpen
  \bibfield  {author} {\bibinfo {author} {\bibfnamefont {A.}~\bibnamefont
  {Wallraff}}, \bibinfo {author} {\bibfnamefont {D.~I.}\ \bibnamefont
  {Schuster}}, \bibinfo {author} {\bibfnamefont {A.}~\bibnamefont {Blais}},
  \bibinfo {author} {\bibfnamefont {L.}~\bibnamefont {Frunzio}}, \bibinfo
  {author} {\bibfnamefont {R.-S.}\ \bibnamefont {Huang}}, \bibinfo {author}
  {\bibfnamefont {J.}~\bibnamefont {Majer}}, \bibinfo {author} {\bibfnamefont
  {S.}~\bibnamefont {Kumar}}, \bibinfo {author} {\bibfnamefont {S.~M.}\
  \bibnamefont {Girvin}},\ and\ \bibinfo {author} {\bibfnamefont {R.~J.}\
  \bibnamefont {Schoelkopf}},\ }\bibfield  {title} {\bibinfo {title} {{Strong
  coupling of a single photon to a superconducting qubit using circuit quantum
  electrodynamics}},\ }\href {https://doi.org/10.1038/nature02851} {\bibfield
  {journal} {\bibinfo  {journal} {Nature}\ }\textbf {\bibinfo {volume} {431}},\
  \bibinfo {pages} {162} (\bibinfo {year} {2004})}\BibitemShut {NoStop}%
\bibitem [{\citenamefont {McHugh}\ \emph {et~al.}(2012)\citenamefont {McHugh},
  \citenamefont {Mazin}, \citenamefont {Serfass}, \citenamefont {Meeker},
  \citenamefont {O'Brien}, \citenamefont {Duan}, \citenamefont {Raffanti},\
  and\ \citenamefont {Werthimer}}]{McHugh2012}%
  \BibitemOpen
  \bibfield  {author} {\bibinfo {author} {\bibfnamefont {S.}~\bibnamefont
  {McHugh}}, \bibinfo {author} {\bibfnamefont {B.~a.}\ \bibnamefont {Mazin}},
  \bibinfo {author} {\bibfnamefont {B.}~\bibnamefont {Serfass}}, \bibinfo
  {author} {\bibfnamefont {S.}~\bibnamefont {Meeker}}, \bibinfo {author}
  {\bibfnamefont {K.}~\bibnamefont {O'Brien}}, \bibinfo {author} {\bibfnamefont
  {R.}~\bibnamefont {Duan}}, \bibinfo {author} {\bibfnamefont {R.}~\bibnamefont
  {Raffanti}},\ and\ \bibinfo {author} {\bibfnamefont {D.}~\bibnamefont
  {Werthimer}},\ }\bibfield  {title} {\bibinfo {title} {{A readout for large
  arrays of microwave kinetic inductance detectors}},\ }\href
  {https://doi.org/10.1063/1.3700812} {\bibfield  {journal} {\bibinfo
  {journal} {Review of Scientific Instruments}\ }\textbf {\bibinfo {volume}
  {83}},\ \bibinfo {pages} {044702} (\bibinfo {year} {2012})},\ \Eprint
  {https://arxiv.org/abs/1203.5861} {arXiv:1203.5861} \BibitemShut {NoStop}%
\bibitem [{\citenamefont {Henderson}\ \emph {et~al.}(2018)\citenamefont
  {Henderson}, \citenamefont {Ahmed}, \citenamefont {Austermann}, \citenamefont
  {Becker}, \citenamefont {Bennett}, \citenamefont {Brown}, \citenamefont
  {Chaudhuri}, \citenamefont {Cho}, \citenamefont {D'Ewart}, \citenamefont
  {Dober}, \citenamefont {Duff}, \citenamefont {Dusatko}, \citenamefont
  {Fatigoni}, \citenamefont {Frisch}, \citenamefont {Gard}, \citenamefont
  {Halpern}, \citenamefont {Hilton}, \citenamefont {Hubmayr}, \citenamefont
  {Irwin}, \citenamefont {Karpel}, \citenamefont {Kernasovskiy}, \citenamefont
  {Kuenstner}, \citenamefont {Kuo}, \citenamefont {Li}, \citenamefont {Mates},
  \citenamefont {Reintsema}, \citenamefont {Smith}, \citenamefont {Ullom},
  \citenamefont {Vale}, \citenamefont {Winkle}, \citenamefont {Vissers},\ and\
  \citenamefont {Yu}}]{Henderson2018}%
  \BibitemOpen
  \bibfield  {author} {\bibinfo {author} {\bibfnamefont {S.~W.}\ \bibnamefont
  {Henderson}}, \bibinfo {author} {\bibfnamefont {Z.}~\bibnamefont {Ahmed}},
  \bibinfo {author} {\bibfnamefont {J.}~\bibnamefont {Austermann}}, \bibinfo
  {author} {\bibfnamefont {D.}~\bibnamefont {Becker}}, \bibinfo {author}
  {\bibfnamefont {D.~A.}\ \bibnamefont {Bennett}}, \bibinfo {author}
  {\bibfnamefont {D.}~\bibnamefont {Brown}}, \bibinfo {author} {\bibfnamefont
  {S.}~\bibnamefont {Chaudhuri}}, \bibinfo {author} {\bibfnamefont {H.-M.~S.}\
  \bibnamefont {Cho}}, \bibinfo {author} {\bibfnamefont {J.~M.}\ \bibnamefont
  {D'Ewart}}, \bibinfo {author} {\bibfnamefont {B.}~\bibnamefont {Dober}},
  \bibinfo {author} {\bibfnamefont {S.~M.}\ \bibnamefont {Duff}}, \bibinfo
  {author} {\bibfnamefont {J.~E.}\ \bibnamefont {Dusatko}}, \bibinfo {author}
  {\bibfnamefont {S.}~\bibnamefont {Fatigoni}}, \bibinfo {author}
  {\bibfnamefont {J.~C.}\ \bibnamefont {Frisch}}, \bibinfo {author}
  {\bibfnamefont {J.~D.}\ \bibnamefont {Gard}}, \bibinfo {author}
  {\bibfnamefont {M.}~\bibnamefont {Halpern}}, \bibinfo {author} {\bibfnamefont
  {G.~C.}\ \bibnamefont {Hilton}}, \bibinfo {author} {\bibfnamefont
  {J.}~\bibnamefont {Hubmayr}}, \bibinfo {author} {\bibfnamefont {K.~D.}\
  \bibnamefont {Irwin}}, \bibinfo {author} {\bibfnamefont {E.~D.}\ \bibnamefont
  {Karpel}}, \bibinfo {author} {\bibfnamefont {S.~S.}\ \bibnamefont
  {Kernasovskiy}}, \bibinfo {author} {\bibfnamefont {S.~E.}\ \bibnamefont
  {Kuenstner}}, \bibinfo {author} {\bibfnamefont {C.-L.}\ \bibnamefont {Kuo}},
  \bibinfo {author} {\bibfnamefont {D.}~\bibnamefont {Li}}, \bibinfo {author}
  {\bibfnamefont {J.~A.~B.}\ \bibnamefont {Mates}}, \bibinfo {author}
  {\bibfnamefont {C.~D.}\ \bibnamefont {Reintsema}}, \bibinfo {author}
  {\bibfnamefont {S.~R.}\ \bibnamefont {Smith}}, \bibinfo {author}
  {\bibfnamefont {J.}~\bibnamefont {Ullom}}, \bibinfo {author} {\bibfnamefont
  {L.~R.}\ \bibnamefont {Vale}}, \bibinfo {author} {\bibfnamefont {D.~D.~V.}\
  \bibnamefont {Winkle}}, \bibinfo {author} {\bibfnamefont {M.}~\bibnamefont
  {Vissers}},\ and\ \bibinfo {author} {\bibfnamefont {C.}~\bibnamefont {Yu}},\
  }\bibfield  {title} {\bibinfo {title} {{Highly-multiplexed microwave SQUID
  readout using the SLAC Microresonator Radio Frequency (SMuRF) electronics for
  future CMB and sub-millimeter surveys}},\ }in\ \href
  {https://doi.org/10.1117/12.2314435} {\emph {\bibinfo {booktitle}
  {Millimeter, Submillimeter, and Far-Infrared Detectors and Instrumentation
  for Astronomy IX}}},\ Vol.\ \bibinfo {volume} {10708},\ \bibinfo {editor}
  {edited by\ \bibinfo {editor} {\bibfnamefont {J.}~\bibnamefont {Zmuidzinas}}\
  and\ \bibinfo {editor} {\bibfnamefont {J.-R.}\ \bibnamefont {Gao}}},\
  \bibinfo {organization} {International Society for Optics and Photonics}\
  (\bibinfo  {publisher} {SPIE},\ \bibinfo {year} {2018})\ pp.\ \bibinfo
  {pages} {170 -- 185}\BibitemShut {NoStop}%
\bibitem [{\citenamefont {{Van Rantwijk}}\ \emph {et~al.}(2016)\citenamefont
  {{Van Rantwijk}}, \citenamefont {Grim}, \citenamefont {{Van Loon}},
  \citenamefont {Yates}, \citenamefont {Baryshev},\ and\ \citenamefont
  {Baselmans}}]{VanRantwijk2016}%
  \BibitemOpen
  \bibfield  {author} {\bibinfo {author} {\bibfnamefont {J.}~\bibnamefont {{Van
  Rantwijk}}}, \bibinfo {author} {\bibfnamefont {M.}~\bibnamefont {Grim}},
  \bibinfo {author} {\bibfnamefont {D.}~\bibnamefont {{Van Loon}}}, \bibinfo
  {author} {\bibfnamefont {S.}~\bibnamefont {Yates}}, \bibinfo {author}
  {\bibfnamefont {A.}~\bibnamefont {Baryshev}},\ and\ \bibinfo {author}
  {\bibfnamefont {J.}~\bibnamefont {Baselmans}},\ }\bibfield  {title} {\bibinfo
  {title} {{Multiplexed Readout for 1000-Pixel Arrays of Microwave Kinetic
  Inductance Detectors}},\ }\href {https://doi.org/10.1109/TMTT.2016.2544303}
  {\bibfield  {journal} {\bibinfo  {journal} {IEEE Trans. Microw. Theory
  Tech.}\ }\textbf {\bibinfo {volume} {64}},\ \bibinfo {pages} {1876} (\bibinfo
  {year} {2016})}\BibitemShut {NoStop}%
\bibitem [{\citenamefont {Mazin}\ \emph
  {et~al.}(2006{\natexlab{b}})\citenamefont {Mazin}, \citenamefont {Day},
  \citenamefont {Irwin}, \citenamefont {Reintsema},\ and\ \citenamefont
  {Zmuidzinas}}]{Mazin2006}%
  \BibitemOpen
  \bibfield  {author} {\bibinfo {author} {\bibfnamefont {B.~A.}\ \bibnamefont
  {Mazin}}, \bibinfo {author} {\bibfnamefont {P.~K.}\ \bibnamefont {Day}},
  \bibinfo {author} {\bibfnamefont {K.~D.}\ \bibnamefont {Irwin}}, \bibinfo
  {author} {\bibfnamefont {C.~D.}\ \bibnamefont {Reintsema}},\ and\ \bibinfo
  {author} {\bibfnamefont {J.}~\bibnamefont {Zmuidzinas}},\ }\bibfield  {title}
  {\bibinfo {title} {{Digital readouts for large microwave low-temperature
  detector arrays}},\ }\href {https://doi.org/10.1016/j.nima.2005.12.208}
  {\bibfield  {journal} {\bibinfo  {journal} {Nucl. Instruments Methods Phys.
  Res. Sect. A Accel. Spectrometers, Detect. Assoc. Equip.}\ }\textbf {\bibinfo
  {volume} {559}},\ \bibinfo {pages} {799} (\bibinfo {year}
  {2006}{\natexlab{b}})}\BibitemShut {NoStop}%
\bibitem [{\citenamefont {Mazin}\ \emph {et~al.}(2013)\citenamefont {Mazin},
  \citenamefont {Meeker}, \citenamefont {Strader}, \citenamefont {Szypryt},
  \citenamefont {Marsden}, \citenamefont {van Eyken}, \citenamefont {Duggan},
  \citenamefont {Walter}, \citenamefont {Ulbricht}, \citenamefont {Johnson},
  \citenamefont {Bumble}, \citenamefont {O'Brien},\ and\ \citenamefont
  {Stoughton}}]{Mazin2013}%
  \BibitemOpen
  \bibfield  {author} {\bibinfo {author} {\bibfnamefont {B.~A.}\ \bibnamefont
  {Mazin}}, \bibinfo {author} {\bibfnamefont {S.~R.}\ \bibnamefont {Meeker}},
  \bibinfo {author} {\bibfnamefont {M.~J.}\ \bibnamefont {Strader}}, \bibinfo
  {author} {\bibfnamefont {P.}~\bibnamefont {Szypryt}}, \bibinfo {author}
  {\bibfnamefont {D.}~\bibnamefont {Marsden}}, \bibinfo {author} {\bibfnamefont
  {J.~C.}\ \bibnamefont {van Eyken}}, \bibinfo {author} {\bibfnamefont {G.~E.}\
  \bibnamefont {Duggan}}, \bibinfo {author} {\bibfnamefont {A.~B.}\
  \bibnamefont {Walter}}, \bibinfo {author} {\bibfnamefont {G.}~\bibnamefont
  {Ulbricht}}, \bibinfo {author} {\bibfnamefont {M.}~\bibnamefont {Johnson}},
  \bibinfo {author} {\bibfnamefont {B.}~\bibnamefont {Bumble}}, \bibinfo
  {author} {\bibfnamefont {K.}~\bibnamefont {O'Brien}},\ and\ \bibinfo {author}
  {\bibfnamefont {C.}~\bibnamefont {Stoughton}},\ }\bibfield  {title} {\bibinfo
  {title} {{ARCONS: A 2024 Pixel Optical through Near-IR Cryogenic Imaging
  Spectrophotometer}},\ }\href {https://doi.org/10.1086/674013} {\bibfield
  {journal} {\bibinfo  {journal} {Publ. Astron. Soc. Pacific}\ }\textbf
  {\bibinfo {volume} {125}},\ \bibinfo {pages} {1348} (\bibinfo {year}
  {2013})},\ \Eprint {https://arxiv.org/abs/1306.4674} {arXiv:1306.4674}
  \BibitemShut {NoStop}%
\bibitem [{Note1()}]{Note1}%
  \BibitemOpen
  \bibinfo {note} {Superconducting Quantum Interference Devices}\BibitemShut
  {NoStop}%
\bibitem [{\citenamefont {Kernasovskiy}\ \emph {et~al.}(2018)\citenamefont
  {Kernasovskiy}, \citenamefont {Kuenstner}, \citenamefont {Karpel},
  \citenamefont {Ahmed}, \citenamefont {{Van Winkle}}, \citenamefont {Smith},
  \citenamefont {Dusatko}, \citenamefont {Frisch}, \citenamefont {Chaudhuri},
  \citenamefont {Cho}, \citenamefont {Dober}, \citenamefont {Henderson},
  \citenamefont {Hilton}, \citenamefont {Hubmayr}, \citenamefont {Irwin},
  \citenamefont {Kuo}, \citenamefont {Li}, \citenamefont {Mates}, \citenamefont
  {Nasr}, \citenamefont {Tantawi}, \citenamefont {Ullom}, \citenamefont
  {Vale},\ and\ \citenamefont {Young}}]{Kernasovskiy2018}%
  \BibitemOpen
  \bibfield  {author} {\bibinfo {author} {\bibfnamefont {S.~A.}\ \bibnamefont
  {Kernasovskiy}}, \bibinfo {author} {\bibfnamefont {S.~E.}\ \bibnamefont
  {Kuenstner}}, \bibinfo {author} {\bibfnamefont {E.}~\bibnamefont {Karpel}},
  \bibinfo {author} {\bibfnamefont {Z.}~\bibnamefont {Ahmed}}, \bibinfo
  {author} {\bibfnamefont {D.~D.}\ \bibnamefont {{Van Winkle}}}, \bibinfo
  {author} {\bibfnamefont {S.}~\bibnamefont {Smith}}, \bibinfo {author}
  {\bibfnamefont {J.}~\bibnamefont {Dusatko}}, \bibinfo {author} {\bibfnamefont
  {J.~C.}\ \bibnamefont {Frisch}}, \bibinfo {author} {\bibfnamefont
  {S.}~\bibnamefont {Chaudhuri}}, \bibinfo {author} {\bibfnamefont {H.~M.}\
  \bibnamefont {Cho}}, \bibinfo {author} {\bibfnamefont {B.~J.}\ \bibnamefont
  {Dober}}, \bibinfo {author} {\bibfnamefont {S.~W.}\ \bibnamefont
  {Henderson}}, \bibinfo {author} {\bibfnamefont {G.~C.}\ \bibnamefont
  {Hilton}}, \bibinfo {author} {\bibfnamefont {J.}~\bibnamefont {Hubmayr}},
  \bibinfo {author} {\bibfnamefont {K.~D.}\ \bibnamefont {Irwin}}, \bibinfo
  {author} {\bibfnamefont {C.~L.}\ \bibnamefont {Kuo}}, \bibinfo {author}
  {\bibfnamefont {D.}~\bibnamefont {Li}}, \bibinfo {author} {\bibfnamefont
  {J.~A.~B.}\ \bibnamefont {Mates}}, \bibinfo {author} {\bibfnamefont
  {M.}~\bibnamefont {Nasr}}, \bibinfo {author} {\bibfnamefont {S.}~\bibnamefont
  {Tantawi}}, \bibinfo {author} {\bibfnamefont {J.}~\bibnamefont {Ullom}},
  \bibinfo {author} {\bibfnamefont {L.}~\bibnamefont {Vale}},\ and\ \bibinfo
  {author} {\bibfnamefont {B.}~\bibnamefont {Young}},\ }\bibfield  {title}
  {\bibinfo {title} {{SLAC Microresonator Radio Frequency (SMuRF) Electronics
  for Read Out of Frequency-Division-Multiplexed Cryogenic Sensors}},\ }\href
  {https://doi.org/10.1007/s10909-018-1981-5} {\bibfield  {journal} {\bibinfo
  {journal} {J. Low Temp. Phys.}\ }\textbf {\bibinfo {volume} {193}},\ \bibinfo
  {pages} {570} (\bibinfo {year} {2018})}\BibitemShut {NoStop}%
\bibitem [{\citenamefont {Palacios-Laloy}\ \emph {et~al.}(2008)\citenamefont
  {Palacios-Laloy}, \citenamefont {Nguyen}, \citenamefont {Mallet},
  \citenamefont {Bertet}, \citenamefont {Vion},\ and\ \citenamefont
  {Esteve}}]{Palacios-Laloy2008a}%
  \BibitemOpen
  \bibfield  {author} {\bibinfo {author} {\bibfnamefont {A.}~\bibnamefont
  {Palacios-Laloy}}, \bibinfo {author} {\bibfnamefont {F.}~\bibnamefont
  {Nguyen}}, \bibinfo {author} {\bibfnamefont {F.}~\bibnamefont {Mallet}},
  \bibinfo {author} {\bibfnamefont {P.}~\bibnamefont {Bertet}}, \bibinfo
  {author} {\bibfnamefont {D.}~\bibnamefont {Vion}},\ and\ \bibinfo {author}
  {\bibfnamefont {D.}~\bibnamefont {Esteve}},\ }\bibfield  {title} {\bibinfo
  {title} {{Tunable resonators for quantum circuits}},\ }\href
  {https://doi.org/10.1007/s10909-008-9774-x} {\bibfield  {journal} {\bibinfo
  {journal} {J. Low Temp. Phys.}\ }\textbf {\bibinfo {volume} {151}},\ \bibinfo
  {pages} {1034} (\bibinfo {year} {2008})},\ \Eprint
  {https://arxiv.org/abs/0712.0221} {arXiv:0712.0221} \BibitemShut {NoStop}%
\bibitem [{\citenamefont {Herrmann}\ \emph {et~al.}(1973)\citenamefont
  {Herrmann}, \citenamefont {Rabiner},\ and\ \citenamefont
  {Chan}}]{Herrmann1973}%
  \BibitemOpen
  \bibfield  {author} {\bibinfo {author} {\bibfnamefont {O.}~\bibnamefont
  {Herrmann}}, \bibinfo {author} {\bibfnamefont {L.~R.}\ \bibnamefont
  {Rabiner}},\ and\ \bibinfo {author} {\bibfnamefont {D.~S.~K.}\ \bibnamefont
  {Chan}},\ }\bibfield  {title} {\bibinfo {title} {{Practical Design Rules for
  Optimum Finite Impulse Response Low-Pass Digital Filters}},\ }\href
  {https://doi.org/10.1002/j.1538-7305.1973.tb01990.x} {\bibfield  {journal}
  {\bibinfo  {journal} {Bell System Technical Journal}\ }\textbf {\bibinfo
  {volume} {52}},\ \bibinfo {pages} {769} (\bibinfo {year} {1973})}\BibitemShut
  {NoStop}%
\bibitem [{\citenamefont {Rabiner}(1973)}]{Rabiner1973}%
  \BibitemOpen
  \bibfield  {author} {\bibinfo {author} {\bibfnamefont {L.}~\bibnamefont
  {Rabiner}},\ }\bibfield  {title} {\bibinfo {title} {{Approximate design
  relationships for low-pass FIR digital filters}},\ }\href
  {https://doi.org/10.1109/TAU.1973.1162510} {\bibfield  {journal} {\bibinfo
  {journal} {IEEE Trans. Audio Electroacoust.}\ }\textbf {\bibinfo {volume}
  {21}},\ \bibinfo {pages} {456} (\bibinfo {year} {1973})}\BibitemShut
  {NoStop}%
\bibitem [{\citenamefont {Crochiere}\ and\ \citenamefont
  {Rabiner}(1983)}]{RonaldE.Crochiere527}%
  \BibitemOpen
  \bibfield  {author} {\bibinfo {author} {\bibfnamefont {R.~E.}\ \bibnamefont
  {Crochiere}}\ and\ \bibinfo {author} {\bibfnamefont {L.~R.}\ \bibnamefont
  {Rabiner}},\ }\href@noop {} {\emph {\bibinfo {title} {Multirate Digital
  Signal Processing}}},\ Vol.~\bibinfo {volume} {1}\ (\bibinfo  {publisher}
  {Prentice-Hall},\ \bibinfo {year} {1983})\BibitemShut {NoStop}%
\bibitem [{\citenamefont {{STEMLab RedPitaya}}()}]{redpitaya}%
  \BibitemOpen
  \bibfield  {author} {\bibinfo {author} {\bibnamefont {{STEMLab RedPitaya}}},\
  }\href {http://www.redpitaya.com/} {\bibinfo {title} {{RedPitaya Open Source
  Instrument}}}\BibitemShut {NoStop}%
\bibitem [{\citenamefont {Bellanger}\ \emph {et~al.}(1976)\citenamefont
  {Bellanger}, \citenamefont {Bonnerot},\ and\ \citenamefont
  {Coudreuse}}]{Bellanger1976}%
  \BibitemOpen
  \bibfield  {author} {\bibinfo {author} {\bibfnamefont {M.}~\bibnamefont
  {Bellanger}}, \bibinfo {author} {\bibfnamefont {G.}~\bibnamefont
  {Bonnerot}},\ and\ \bibinfo {author} {\bibfnamefont {M.}~\bibnamefont
  {Coudreuse}},\ }\bibfield  {title} {\bibinfo {title} {{Digital filtering by
  polyphase network:Application to sample-rate alteration and filter banks}},\
  }\href {https://doi.org/10.1109/TASSP.1976.1162788} {\bibfield  {journal}
  {\bibinfo  {journal} {IEEE Trans. Acoust.}\ }\textbf {\bibinfo {volume}
  {24}},\ \bibinfo {pages} {109} (\bibinfo {year} {1976})}\BibitemShut
  {NoStop}%
\bibitem [{\citenamefont {Vaidyanathan}(1990)}]{Vaidyanathan1990a}%
  \BibitemOpen
  \bibfield  {author} {\bibinfo {author} {\bibfnamefont {P.~P.}\ \bibnamefont
  {Vaidyanathan}},\ }\bibfield  {title} {\bibinfo {title} {{Multirate Digital
  Filters, Filter Banks, Polyphase Networks, and Applications: A Tutorial}},\
  }\href {https://doi.org/10.1109/5.52200} {\bibfield  {journal} {\bibinfo
  {journal} {Proc. IEEE}\ }\textbf {\bibinfo {volume} {78}},\ \bibinfo {pages}
  {56} (\bibinfo {year} {1990})}\BibitemShut {NoStop}%
\bibitem [{\citenamefont {Vetterli}(1987)}]{Vetterli1987}%
  \BibitemOpen
  \bibfield  {author} {\bibinfo {author} {\bibfnamefont {M.}~\bibnamefont
  {Vetterli}},\ }\bibfield  {title} {\bibinfo {title} {{A Theory of Multirate
  Filter Banks}},\ }\href {https://doi.org/10.1109/TASSP.1987.1165137}
  {\bibfield  {journal} {\bibinfo  {journal} {IEEE Trans. Acoust.}\ }\textbf
  {\bibinfo {volume} {35}},\ \bibinfo {pages} {356} (\bibinfo {year}
  {1987})}\BibitemShut {NoStop}%
\bibitem [{\citenamefont {Fliege}(1994)}]{NorbertFliege633}%
  \BibitemOpen
  \bibfield  {author} {\bibinfo {author} {\bibfnamefont {N.}~\bibnamefont
  {Fliege}},\ }\href@noop {} {\emph {\bibinfo {title} {Multirate Digital Signal
  Processing: Multirate Systems, Filter Banks, Wavelets}}},\ Vol.~\bibinfo
  {volume} {1}\ (\bibinfo  {publisher} {Wiley},\ \bibinfo {year}
  {1994})\BibitemShut {NoStop}%
\bibitem [{\citenamefont {Lyons}(1997)}]{RichardG.Lyons85}%
  \BibitemOpen
  \bibfield  {author} {\bibinfo {author} {\bibfnamefont {R.~G.}\ \bibnamefont
  {Lyons}},\ }\href@noop {} {\emph {\bibinfo {title} {Understanding Digital
  Signal Processing}}},\ Vol.~\bibinfo {volume} {1}\ (\bibinfo  {publisher}
  {Addison Wesley Publishing Company},\ \bibinfo {year} {1997})\BibitemShut
  {NoStop}%
\bibitem [{\citenamefont {Wei{\ss}}\ \emph
  {et~al.}(1998{\natexlab{a}})\citenamefont {Wei{\ss}}, \citenamefont
  {Harteneck},\ and\ \citenamefont {Stewart}}]{Weib1998b}%
  \BibitemOpen
  \bibfield  {author} {\bibinfo {author} {\bibfnamefont {S.}~\bibnamefont
  {Wei{\ss}}}, \bibinfo {author} {\bibfnamefont {M.}~\bibnamefont
  {Harteneck}},\ and\ \bibinfo {author} {\bibfnamefont {R.~W.}\ \bibnamefont
  {Stewart}},\ }\bibfield  {title} {\bibinfo {title} {{Design and Efficient
  Implementation of Oversampled {\{}GDFT{\}} Filter Banks for Subband Adaptive
  Filtering}},\ }\href {http://eprints.soton.ac.uk/id/eprint/251914} {\bibfield
   {journal} {\bibinfo  {journal} {Digest IEE Colloquium on Digital Filters: an
  Enabling Technology}\ ,\ \bibinfo {pages} {12/1}} (\bibinfo {year}
  {1998}{\natexlab{a}})}\BibitemShut {NoStop}%
\bibitem [{\citenamefont {Wei{\ss}}\ \emph
  {et~al.}(1998{\natexlab{b}})\citenamefont {Wei{\ss}}, \citenamefont
  {Harteneck},\ and\ \citenamefont {Stewart}}]{Weib1998c}%
  \BibitemOpen
  \bibfield  {author} {\bibinfo {author} {\bibfnamefont {S.}~\bibnamefont
  {Wei{\ss}}}, \bibinfo {author} {\bibfnamefont {M.}~\bibnamefont
  {Harteneck}},\ and\ \bibinfo {author} {\bibfnamefont {R.~W.}\ \bibnamefont
  {Stewart}},\ }\bibfield  {title} {\bibinfo {title} {{On implementation and
  design of filter banks for subband adaptive systems}},\ }\href@noop {}
  {\bibfield  {journal} {\bibinfo  {journal} {IEE Colloquium (Digest)}\ }
  (\bibinfo {year} {1998}{\natexlab{b}})}\BibitemShut {NoStop}%
\bibitem [{\citenamefont {Peng}\ \emph {et~al.}(2013)\citenamefont {Peng},
  \citenamefont {Liu},\ and\ \citenamefont {Mandic}}]{Peng2013}%
  \BibitemOpen
  \bibfield  {author} {\bibinfo {author} {\bibfnamefont {B.}~\bibnamefont
  {Peng}}, \bibinfo {author} {\bibfnamefont {W.}~\bibnamefont {Liu}},\ and\
  \bibinfo {author} {\bibfnamefont {D.~P.}\ \bibnamefont {Mandic}},\ }\bibfield
   {title} {\bibinfo {title} {{Design of oversampled generalised discrete
  Fourier transform filter banks for application to subband‐based blind
  source separation}},\ }\href {https://doi.org/10.1049/iet-spr.2012.0361}
  {\bibfield  {journal} {\bibinfo  {journal} {IET Signal Process.}\ }\textbf
  {\bibinfo {volume} {7}},\ \bibinfo {pages} {843} (\bibinfo {year}
  {2013})}\BibitemShut {NoStop}%
\bibitem [{\citenamefont {Proakis}\ and\ \citenamefont
  {Manolakis}(2006)}]{JohnG.Proakis663}%
  \BibitemOpen
  \bibfield  {author} {\bibinfo {author} {\bibfnamefont {J.~G.}\ \bibnamefont
  {Proakis}}\ and\ \bibinfo {author} {\bibfnamefont {D.~G.}\ \bibnamefont
  {Manolakis}},\ }\href@noop {} {\emph {\bibinfo {title} {Digital Signal
  Processing}}},\ Vol.~\bibinfo {volume} {1}\ (\bibinfo  {publisher}
  {Pearson},\ \bibinfo {year} {2006})\BibitemShut {NoStop}%
\bibitem [{\citenamefont {Bellanger}\ and\ \citenamefont
  {Daguet}(1974)}]{Bellanger1974}%
  \BibitemOpen
  \bibfield  {author} {\bibinfo {author} {\bibfnamefont {M.}~\bibnamefont
  {Bellanger}}\ and\ \bibinfo {author} {\bibfnamefont {J.}~\bibnamefont
  {Daguet}},\ }\bibfield  {title} {\bibinfo {title} {{TDM-FDM Transmultiplexer:
  Digital Polyphase and FFT}},\ }\href
  {https://doi.org/10.1109/TCOM.1974.1092391} {\bibfield  {journal} {\bibinfo
  {journal} {IEEE Trans. Commun.}\ }\textbf {\bibinfo {volume} {22}},\ \bibinfo
  {pages} {1199} (\bibinfo {year} {1974})}\BibitemShut {NoStop}%
\bibitem [{Note2()}]{Note2}%
  \BibitemOpen
  \bibinfo {note} {Using the \protect \emph {type-3} polyphase representation.
  For details, see~\cite {NorbertFliege633} and \cite
  {RonaldE.Crochiere527}.}\BibitemShut {Stop}%
\bibitem [{\citenamefont {{Zynq-7000 SoC Data Sheet: Overview (DS190),
  v1.11.1}}(2018)}]{ds190}%
  \BibitemOpen
  \bibfield  {author} {\bibinfo {author} {\bibnamefont {{Zynq-7000 SoC Data
  Sheet: Overview (DS190), v1.11.1}}},\ }\href
  {https://www.xilinx.com/support/documentation/data_sheets/ds190-Zynq-7000-Overview.pdf}
  {} (\bibinfo {year} {2018})\BibitemShut {NoStop}%
\bibitem [{\citenamefont {{LogiCORE IP FIR Compiler Product Guide (PG149),
  v7.2}}(2020)}]{pg149}%
  \BibitemOpen
  \bibfield  {author} {\bibinfo {author} {\bibnamefont {{LogiCORE IP FIR
  Compiler Product Guide (PG149), v7.2}}},\ }\bibfield  {title} {\bibinfo
  {title} {{LogiCORE IP FIR Compiler Product Guide (PG149), v7.2}},\ }\href
  {https://www.xilinx.com/support/documentation/ip_documentation/fir_compiler/v7_2/pg149-fir-compiler.pdf}
  {\ \textbf {\bibinfo {volume} {June}} (\bibinfo {year} {2020})}\BibitemShut
  {NoStop}%
\bibitem [{\citenamefont {{LogiCORE IP FFT Product Guide (PG109),
  v9.1}}(2020)}]{pg109}%
  \BibitemOpen
  \bibfield  {author} {\bibinfo {author} {\bibnamefont {{LogiCORE IP FFT
  Product Guide (PG109), v9.1}}},\ }\bibfield  {title} {\bibinfo {title}
  {{LogiCORE IP FFT Product Guide (PG109), v9.1}},\ }\href
  {https://www.xilinx.com/support/documentation/ip_documentation/xfft/v9_1/pg109-xfft.pdf}
  {\ \textbf {\bibinfo {volume} {June}} (\bibinfo {year} {2020})}\BibitemShut
  {NoStop}%
\bibitem [{\citenamefont {Berner}\ and\ \citenamefont
  {Leon}(1999)}]{Berner1999}%
  \BibitemOpen
  \bibfield  {author} {\bibinfo {author} {\bibfnamefont {S.}~\bibnamefont
  {Berner}}\ and\ \bibinfo {author} {\bibfnamefont {P.~L.~D.}\ \bibnamefont
  {Leon}},\ }\bibfield  {title} {\bibinfo {title} {{FPGA-Based Filterbank
  Implementation for Parallel Digital Signal Processing}},\ }\href
  {http://www.ece.nmsu.edu/$\sim$pdeleon/Research/Publications/NASA_VLSI_Symposium_1999.pdf}
  {\bibfield  {journal} {\bibinfo  {journal} {Proc. NASA Sym. VLSI Des.}\ }
  (\bibinfo {year} {1999})}\BibitemShut {NoStop}%
\bibitem [{\citenamefont {Satorius}\ \emph {et~al.}(2006)\citenamefont
  {Satorius}, \citenamefont {Wu}, \citenamefont {LaRocca},\ and\ \citenamefont
  {Kosinski}}]{Satorius2006}%
  \BibitemOpen
  \bibfield  {author} {\bibinfo {author} {\bibfnamefont {E.}~\bibnamefont
  {Satorius}}, \bibinfo {author} {\bibfnamefont {Y.~W.}\ \bibnamefont {Wu}},
  \bibinfo {author} {\bibfnamefont {B.}~\bibnamefont {LaRocca}},\ and\ \bibinfo
  {author} {\bibfnamefont {J.}~\bibnamefont {Kosinski}},\ }\bibfield  {title}
  {\bibinfo {title} {{Implementation of polyphase channelizers for multirate
  signal analysis}},\ }in\ \href {https://doi.org/10.1109/ACSSC.2006.354939}
  {\emph {\bibinfo {booktitle} {Conf. Rec. - Asilomar Conf. Signals, Syst.
  Comput.}}}\ (\bibinfo  {publisher} {IEEE},\ \bibinfo {year} {2006})\ pp.\
  \bibinfo {pages} {1170--1174}\BibitemShut {NoStop}%
\bibitem [{\citenamefont {{De Leon}}(1998)}]{DeLeon1998}%
  \BibitemOpen
  \bibfield  {author} {\bibinfo {author} {\bibfnamefont {P.~L.}\ \bibnamefont
  {{De Leon}}},\ }\bibfield  {title} {\bibinfo {title} {{On the Use of Filter
  Banks for Parallel Digital Signal Processing}},\ }\href
  {http://www.ece.nmsu.edu/{~}pdeleon/Research/Publications/NASA{\_}VLSI{\_}Symposium{\_}1998.pdf}
  {\bibfield  {journal} {\bibinfo  {journal} {Proc. NASA Sym. on VLSI Design}\
  } (\bibinfo {year} {1998})}\BibitemShut {NoStop}%
\bibitem [{\citenamefont {Parks}\ and\ \citenamefont
  {McClellan}(1972)}]{Parks1972}%
  \BibitemOpen
  \bibfield  {author} {\bibinfo {author} {\bibfnamefont {T.}~\bibnamefont
  {Parks}}\ and\ \bibinfo {author} {\bibfnamefont {J.}~\bibnamefont
  {McClellan}},\ }\bibfield  {title} {\bibinfo {title} {{Chebyshev
  Approximation for Nonrecursive Digital Filters with Linear Phase}},\ }\href
  {https://doi.org/10.1109/TCT.1972.1083419} {\bibfield  {journal} {\bibinfo
  {journal} {IEEE Transactions on Circuit Theory}\ }\textbf {\bibinfo {volume}
  {19}},\ \bibinfo {pages} {189} (\bibinfo {year} {1972})}\BibitemShut
  {NoStop}%
\end{thebibliography}%

	\end{document}